\documentstyle[preprint,eqsecnum,aps]{revtex}

\def\px{\psi(x)}
\def\C{{\cal C}}
\def\B{{\cal B}}
\def\F{{\cal F}}
\def\P{{\cal P}}
\def\H{{\cal H}}
\def\W{{\cal W}}
\def\M{{\cal M}}
\def\pa{\partial}
\def\a{\alpha}
\def\l{\lambda}
\def\g{\gamma}
\def\si{\sigma}
\def\te{\tilde\eta}
\def\ha{\frac{1}{2}}
\def\ve{\varepsilon}
\def\px{\psi(x)}
\def\C{{\cal C}}
\def\P{{\cal P}}
\def\H{{\cal H}}
\def\pa{\partial}
\def\ual{\underline{\alpha}}
\def\Lk{{\Lambda^{-1}_k}}
\def\Lp{\Lambda^{-1}_p}
\def\hps{\hat\psi^{(s)}}
\def\hpss{{\hps}_{s_3}}
\def\Us{U^{(s)}}
\def\Vpm{V^\pm_m}
\def\Mpm{{\cal M}^\pm_m}
\def\Sipm{\Sigma^\pm_m (x)}
\def\Vpmm{V^\pm_m\times\C_2}
\def\usp{u(s,s'_3)}
\def\uss{u(s,s_3)}
\def\mn{\mu\nu}
\def\abg{\alpha,\beta,\gamma}
\def\ha{\frac{1}{2}}
\def\tes{{\te}^{(s)}_{q,p}}
\def\Hms{\H_{[m,s]}}
\def\bfPsi{{\bf\Psi}}
\def\psis{\psi^{(s)}(q,p;\a)}
\def\psisx{\psi^{(s)}_x(q,p;\a)}
\def\Fisi{{\bf\Phi}^{\bar\sigma (x)}_{q,p}}
\def\bPsi{{\bf\Psi}^{(s)}_x(\alpha )}
\def\pss{\psi_{s_3}^{(s)}}
\def\Hs{\H^{(s)}_{\te}}
\def\tohoch^#1{\mathrel{\mathop{\longrightarrow}\limits^{#1}}}
\def\ua{\underline{a}}
 
\begin{document}
\draft
\preprint{MPI-PhT/96-52}
\title{Geometro-Stochastically Quantized Fields \\
               with Internal Spin Variables}
\author{W. Drechsler}
\address{Max-Planck-Institut f\"ur Physik\\
         F\"ohringer Ring 6, 80805 M\"unchen, Germany}
\maketitle
\begin{abstract}
\renewcommand{\baselinestretch}{1.0}
\small\normalsize
The use of internal variables for the description of relativistic
particles with arbitrary mass and spin in terms of scalar functions
is reviewed and applied to the stochastic phase space formulation
of quantum mechanics. Following Bacry and Kihlberg a four-dimensional
internal spin space $\bar S$ is chosen possessing an invariant
measure and being able to represent integer as well as half integer spins.
$\bar S$ is a homogeneous space of the  group $SL(2,\C)$ parametrized
in terms of spinors $\a\in\C_2$ and their complex conjugates $\bar\a$.
The generalized scalar quantum mechanical wave functions may be reduced to
yield irreducible components of definite physical mass and spin $[m,s]$,
with $m\ge 0$ and $s=0,\ha,1,\frac{3}{2}\dots$ , with spin described
in terms of the usual $(2s+1)$-component fields. Viewed from the internal
space description of spin this reduction amounts to a restriction
of the variable $\a$ to a compact subspace of $\bar S$, i.e. a
``spin shell" $S^2_{r=2s}$ of radius $r=2s$ in $\C_2$. This formulation
of single particles or single antiparticles of type $[m,s]$ is then
used to study the geometro-stochastic (i.e. quantum) propagation of
amplitudes for arbitrary spin on a curved background space-time
possessing a metric and axial vector torsion treated as external
fields. A Poincar\'e gauge covariant path integral-like representation
for the probability amplitude (generalized wave function) of a particle
with arbitrary spin is derived satisfying a second order wave equation
on the Hilbert bundle constructed over curved space-time. The 
implications for the stochastic nature of polarization effects in
the presence of gravitation are pointed out and the extension to
Fock bundles of bosonic and fermionic type is briefly mentioned.
\end{abstract}
\newpage
 
\section{Introduction}
Spin appeared in physics as a typical property of quantum
mechanical states determining the multiplet structure of atomic
spectra. The concept of a nonrelativistic quantum mechanical wave
function had to be broadened to be able to account for the presence
of spin yielding thereby a unified description for orbital as
well as spin motion formulated with the help of group theoretical
methods\cite{Weyl}, or, more precisely, treated
in terms of the representation
theory of the rotation group $SO(3)$ \cite{Wigner,Ed}.
Extending this
nonrelativistic theory to a formulation in accord with special
relativity, within a Lorentz and translation invariant formalism
for free particles, leads to Wigner's \cite{Wig}
identification of elementary
particles observed in nature
with the unitary irreducible representations (UIR) of the
Poincar\'e group, $\P=ISO(3,1)$, characterized by mass and spin
$[m,s]$, with real $m\geq 0$ and $s=0,\ha,1,\frac{3}{2},
2\ldots$.
 
Relativistic particles with definite mass and spin are in this
scheme described in terms of multicomponent fields,
$\psi_a (x); a=1\ldots n$, $x=(x^\mu;\mu=0,1,2,3)$,
defined over Minkowski space-time
$M_4$, which transform as vectors under $\P$ or, in the half integer
spin cases, as spinors under the universal covering goup
$\bar\P=\overline{ISO (3,1)}=T_4 \otimes_s  SL(2,\C)$, where
$\otimes_s$ denotes the semidirect product.
The interaction between various different fields is then usually
formulated as a Lorentz invariant coupling of these
multicomponent fields, for example, like $e\bar\psi(x)\g_\mu\px
A^\mu (x)$ in the case of QED or like
$g\bar\psi_N(x)\g_5\vec\tau\psi_N(x)\vec\phi (x)$ as, for example,
for the pseudoscalar pion nucleon interaction\cite{text}.
Writing the interaction between different elementary particle
fields in this manner freezes the spin content of the fields to their
free field values and thus forces the spin degrees of freedom to play
an undynamical role in the theory.
 
When the Regge theory of strong interaction was en vogue in particle
physics it was observed from the data on high energy collisions that
the effective spin of the exchanged particle mediating the strong
interaction, for example the $\rho$-meson or the pion,
depended on the momentum transfer $\bar t$ between the collision partners
taking part in the process and was {\it not} a constant given by the
fixed spin value of the free field. There was a so-called trajectory
relation involved connecting the scattering states for $\bar t<0$ and
continuously variable effective spin with a family of
Regge recurrences for $\bar t=m^2_s>0$ and discrete physical integer or
half integer spin values $s$ of certain observed resonant states i.e.\
excited states of strong interaction. This showed that in going
``off mass shell'' with the invariant
energy or momentum transfer variable
of an analytic $S$-matrix element one had in strong interaction
physics also to go ``off spin shell'' and analytically continue
in the spin variable, i.e.\ one had to allow for a dynamical
role of spin \cite{Lur}.
 
In this context one intended in the sixties to replace the
elementary fields for particles of definite mass and spin by so-called
spin-tower-fields with built-in trajectory relation between mass and
spin (compare, for example, Bacry and Nuyts\cite{BN}). At the same time
the question was asked whether it would be appropriate to represent
a particle with spin not by a vector- or spinor-valued
function over Minkowski space-time with fixed number of components
but by a scalar function defined over a higher dimensional space,
in particular, a homogeneous space of the underlying kinematic symmetry
group $\P$, i.e.\ to consider fields in particle physics as
scalar-valued functions defined on $\P/H$ where $H$ is a closed
subgroup of $\P$\cite{Lur,Finkel,BK}.
In order to consider homogeneous spaces of the Poincar\'e group
which contain Minkowski space-time, i.e.\
being of the type $M_4\times S$, with $S$ playing the role of a
spin space, one regards $H$ to be a subgroup of the Lorentz
group contained in $\P$ yielding thereby for the space $S$ a homogeneous
space of the Lorentz group.
 
All the homogeneous spaces of the Poincar\'e
group of this type were listed
by Finkelstein\cite{Finkel} and by Bacry and Kihlberg\cite{BK}, and
the existence of an invariant measure on $\P/H$
as well as the suitability
of these spaces for the description of half integer spins were
investigated. The authors of Ref.~\cite{BK} came to the conclusion
that the lowest dimensional homogeneous space with invariant measure
suitable for the description of half integer as well as integer spins
is eight-dimensional with four internal variables for the
representation of spin, $S$=``[4]''
in Finkelstein's notation,
yielding thereby a generalized scalar wave function, $\psi(X)$,
for the description of a particle with spin, with $X=(x^\mu,y^i)\in
\P/H$ where $x^\mu\in M_4$ and $y^i\in S$, $i=1,2,3,4$.
 
If one wants to have fields with a fixed mass value $m$ and a definite
spin $s$ it is required that the field $\psi (X)$ takes sharp values
for the two Casimir operators of the Poincar\'e group, i.e.
\begin{eqnarray}
\hat P_\mu \hat P^\mu \psi(X) & = &\phantom{-}m^2\psi(X),\label{11}\\
\hat W_\mu\hat W^\mu\psi(X)  & = & -m^2s(s+1)\psi (X),   \label{12}
\end{eqnarray}
with $\hat P_\mu$ denoting the momentum operator and with
\begin{equation}
\hat W_\mu=\frac{1}{2} \ve_{\mu\nu\rho\l}\hat P^\nu\hat S^{\rho\l}
\label{13}
\end{equation}
being the Pauli-Lubanski operator, where
$\hat S^{\rho\l}=-\hat S^{\l\rho}$
is a set of spin operators satisfying the Lie algebra of
$SO(3,1)$ which are
expressed as differential operators in the additional internal spin
variables, $y^i$, with the coordinates $y^i$ parametrizing the space
$S=SO(3,1)/H$.  [$SO(3,1)$ is used here
to denote the proper orthochronous
Lorentz group $O(3,1)^{++}$.]
 
However, it was pointed out by Bacry and Kihlberg \cite{BK}
that in order
to reduce the description completely to an irreducible one
in the Wigner sense two additional conditions on the scalar
wave functions for the quantized description of spin had to
be introduced. These could most easily be expressed by using a
two-dimensional spinor formulation of the internal spin space
\begin{equation}
\bar S=SL(2,\C)/\bar H
\label{14}
\end{equation}
where $\bar H$ (being the universal covering group of $H$) is a
subgroup of $SL(2,\C)$ determining the space
defined by Eq.(\ref{14}) \cite{space}.
The space $\bar S$ could thus be parametrized by spinor variables
$\a^A, A=1,2$, given by \cite{BK}
\begin{eqnarray}
\a^1 & = &e^{\ha t}e^{i\ha\psi}\cos\ha\theta~e^{i\ha\varphi}\/,
\nonumber\\
\a^2 & = &e^{\ha t}e^{i\ha\psi}\sin\ha\theta e^{-i\ha\varphi}\/,
\label{15}
\end{eqnarray}
and their complex conjugates 
$\bar\alpha^{{\dot A}}; \dot A=\dot 1,\dot 2$.
Here $-\infty < t <\infty$; $0\leq \theta \leq \pi$,
$0\leq\varphi <2\pi$, and $-2\pi \leq \psi <2\pi$.
This shows that the internal spin space with the four real variables
$t,\psi,\theta, \varphi$ and the Lorentz invariant measure
\begin{equation}
d\mu_s=e^{2t}dt d\psi d\cos\theta d\varphi
\label{16}
\end{equation}
is isomorphic to a two-dimensional complex space $\C_2$ with
measure $d\alpha d\bar\alpha = d\alpha^1d\alpha^2d\bar\alpha^{\dot 1}
d\bar\alpha^{{\dot 2}}$ being invariant because of the unimodularity of
$SL(2,\C)$.
From this point of view one would represent the
wave function for a relativistic particle with arbitrary spin as a
scalar function $\psi (x,\alpha,\bar\alpha)$ with
$x=(x^\mu;\mu=0,1,2,3)$, denoting a point in Minkowski space,
and with $\alpha=(\alpha^A; A=1,2)$ and its complex conjugate
$\bar\a=(\bar\a^{\dot A}; \dot A=\dot 1,\dot 2)$
denoting a point in the internal spin space $\C_2$ and its complex
conjugate, respectively. $\psi(x,\a,\bar\a)$
would have to satisfy the Klein-Gordon equation in $x$
which is essentially Eq.~(\ref{11}) with the velocity of light
taken to be $c=1$.
 
In parentheses we would like to remark already at this place that due
to the impossibility of localizing a relativistic particle in Minkowski
space in terms of projector-valued (PV) measures providing a
system of imprimitivity of the Poincar\'e group in Mackey's sense
on the Hilbert space of states \cite{alpha,beta}, we shall
present below a {\it stochastic phase space} description of
relativistic particles, as advocated strongly by Prugove\v{c}ki
\cite{prug1,prug2,Ali}, and construct a system of covariance of the
Poincar\'e group in terms of {\it positive operator-valued} (POV)
measures on a Hilbert space for particles with arbitrary
spin. This yields a stochastic phase space description of
relativistic particles as proposed by Prugove\v{c}ki which is
extended in this paper to nonzero spin by using a homogeneous space
description of the Poincar\'e group for the spin degrees of freedom --
or rather a $\C_2$-description as mentioned above -- leading to a
fully covariant formalism for the kinematics and localization
properties of free relativistic particles with definite mass
and spin in terms of scalar functions. At a later stage of this
investigation we shall study the implications of the internal space
description of spin for the coupling of fields describing interactions
among several particles with nonzero (dynamical) spin, i.e.
couple several general spin fields together in using the internal
spinorial variables introduced and investigated in the present paper.
 
Returning now to the scalar wave function $\psi=\psi (x,\a,\bar\a)$
with internal spinorial variables $\a$ and $\bar\a$ transforming under
the fundamental representations $D^{(\ha,0)}(\Lambda)$ and
$D^{(0,\ha)}(\Lambda)$ of $SL(2,\C)$, respectively, one demands,
following Bacry and Kihlberg \cite{BK}, for the states of
definite spin $s$ that the following
two constraining equations are satisfied:
\begin{equation}
\hat D\psi =\a^A\frac{\partial\psi}{\partial\a^A}=2s\psi\/,
\label{17}
\end{equation}
and
\begin{equation}
\frac{\partial}{\pa\bar\a^{\dot A}}\psi =0\/.\label{18}
\end{equation}
If Eqs.\ (\ref{17}) and (\ref{18}) are obeyed, $\psi$ is a homogeneous
polynomial of degree $2 s$ in the undotted spinor variables $\a^1$
and $\a^2$ fixing the spin to the integer or half integer value $s$ with
no dependence of $\psi$ on the dotted spinor variables $\bar\a^{\dot 1}$
and $\bar\a^{\dot 2}$. We call Eq.~(\ref{17}) [with the summation
convention used for the spinor indices]
the {\it homogeneity condition} reducing the wave function $\psi$ for
arbitrary spin to a particular spin value $s$; and we call
Eq.~(\ref{18}) the {\it holomorphicity condition}
yielding thus a spin description
in terms of holomorphic functions of the variables 
$\alpha\in \C_2$.
 
It follows from Ref.\cite{BK} that if Eqs.~(\ref{17}), (\ref{18})
and (\ref{11}) are satisfied by $\psi$ also the
Casimir operator appearing on the l.-h.side
of Eq.~(\ref{12}) possesses a sharp eigenvalue for $\psi$ given by
$-m^2s(s+1)$ and the description reduces to an irreducible one in
the Wigner sense.
 
It is interesting to remark that Eqs.~(\ref{17}) and (\ref{18}) are
instructive also from another point of view. In the course of
investigating the geometric quantization of constrained Hamiltonian
systems describing particles with nonzero spin one proceeds by
giving a classical phase space description of spin in extending
the symplectic geometry, i.e.\ the phase space geometry, to include
the spin degrees of freedom which -- after quantization -- yield
discrete integer values for $2s$ and transform irreducibly under
$SU(2)$, in the nonrelativistic case, or under $SL(2,\C)$, in
the relativistic case (compare N.\ Woodhouse \cite{Wood}).
Classically, the subspace of $\C_2$ appropriate for the
description of a definite spin $s$ is the two-sphere, $S^2_{r=2s}=
S^3_{r=2s}/U(1)$ obtained as a factorization by $U(1)$ of the
three-sphere $S^3_{r=2s}$ of radius $r=2s$, i.e.\ the ``spin
shell'' given by \begin{equation}
p_{A\dot A}\a^A\bar\a^{{\dot A}}=\bar r = mcr~,
\label{19}
\end{equation}
where $p_{A\dot A}=(1p^0+\vec\sigma\vec p)_{A\dot A}$, with the
Pauli matrices $\vec\sigma =(\sigma_1,\sigma_2,\sigma_3)$,
is the spinor form of the 4-momentum $p^\mu=(p^0,\vec p)$.
Here the $U(1)$ degree of freedom
$(\a^A,\bar\a^{\dot A})\to(e^{i\chi}\a^A,e^{-i\chi}\bar\a^{\dot A})$,
with real $\chi$, defines an equivalence class for
which Eq.~(\ref{19}) remains unchanged. After quantization this yields
through Eqs.~(\ref{17}) and (\ref{18}) a description in terms of a
momentum and $\alpha$-dependent {\it reduced} wave function
$\tilde\psi^{(s)}(p,\a)$, with $\a$ defined on a sphere $S^2_{r=2s}$
of radius $r=2s$, with integer $r$, contained in $\C_2$.
The two-spheres $S^2_{r=2s}$ of non-zero integer radius
$r=2s$ define an {\it integral orbit space} representing the
coadjoint orbits of the rotation group in Kirillov's terminology
\cite{Kir}.
 
It appears from Eqs.~(\ref{17}) and (\ref{18}) that only {\it half}
of the phase space variables $\a^1,\a^2$ and $\bar\a^{\dot 1},
\bar\a^{\dot 2}$ are relevant for a
quantized description of physical spin and that
the spin space for free noninteracting
particles is essentially a two-sphere.
A two-sphere $S^2$ may be regarded, on the one hand, as the
homogeneous space $SU(2)/U(1)$ or, on the other hand, as the
homogeneous space $SL(2,\C)/\tilde P$ where $\tilde P$ is the subgroup of
$SL(2,\C)$ of complex triangular matrices of the form
$\pmatrix{\rho&0\cr\eta&\rho^{-1}}$ with
$\rho,\eta\in\C$ (see \cite{Evans}). The universal covering group
$SL(2,\C)$ of the Lorentz group acts transitively on the
two-sphere in this latter form; it acts on $S^2=SU(2)/U(1)$ through the
Wigner rotation $\pm\bar R(p,\Lambda)\in SU(2)$ related to the
Lorentz transformation $\Lambda\in SO(3,1)$ carrying a momentum
eigenstate with momentum $p$ into one with momentum $\Lambda p$, i.e.
\begin{equation}
\Lambda =\Lambda_{\Lambda p}R(p,\Lambda)\Lambda^{-1}_p
\label{110}
\end{equation}
where $\Lambda_p$ is the boost, $p=\Lambda_p\!\buildrel o\over p$,
taking the rest momentum $\buildrel o\over p=(mc,\vec 0)$ into
$p=(p^o,\vec p)$, and $\pm\bar R(p,\Lambda)$ above is the element
of $SU(2)$ corresponding to the $SO(3)$ rotation
$R(p,\Lambda)$ in (\ref{110}).
 
As viewed from the original space $\C_2$, the reduction involved
in the quantized description of arbitrary spin in terms of a function
$\tilde\psi(p,\alpha,\bar\alpha)$ depending $\a$ and its complex
conjugate to a function
$\tilde\psi^{(s)}(p,\a)$ for definite spin defined on a two-sphere
in $\C_2$ may be regarded
also in the following way. One may view the space $\C_2$ as a
$GL(1,\C)$-bundle over $S^2$,
\begin{equation}
\C_2=P(S^2, GL(1,\C))
\label{111}
\end{equation}
with $GL(1,\C)=\C^\star =\C\backslash\{0\}$ being isomorphic to
the complex numbers without the origin. For $\lambda\in\C^\star$
the $GL(1,\C)$ transformations give rise to an equivalence relation
provided by $(\a^A,\bar\a^{\dot A})\sim (\l\a^A,\bar\l\bar\a^{\dot A})$.
The two-sphere $S^2$ may thus be regarded
as the space $\C_2$ modulo this equivalence relation describing
dilatations by $\l$ \cite{Popov}.
Hence, the reduction originating from the
imposition of Eqs.~(\ref{17}) and (\ref{18})
implies a corresponding reduction
given by a projection in the principal bundle (\ref{111}) from the
bundle space to its base.
 
The key observation in the context of the present paper, however,
is that in contradistinction to the vector-type representations
for spin appearing in Refs.\cite{AliP} and \cite{Ali2}
based on Wigner rotations, spin
can, indeed, be given a formulation in terms of scalar functions
defined on a four-dimensional homogeneous space of the Lorentz
group possessing an invariant measure with the internal spinorial
variables transforming under $SL(2,\C)$. This description reduces, as
mentioned, to an irreducible one corresponding to a definite mass $m$
and spin $s$ for free physical particles
if, besides (\ref{11}), the constraints (\ref{17}) and (\ref{18}) are
required to be satisfied. In this case results similar to those of
Ali and Prugove\v{c}ki\cite{AliP} are obtained, where, in our presentation, a
joint description for integer as well as half integer spins is given.
This is due to the fact that even in the
reduced case the internal spin variables may be considered to
transform under $SL(2,\C)$ although we know that the internal
spin space, in fact, reduces to a subspace of $\C_2$, i.e.\ to a
two-sphere (a compact space) and the transformation group may
be considered to reduce to the group of Wigner rotations, i.e.\
to the transformations of the compact subgroup
$SO(3)$ of the Lorentz group or
its covering group $SU(2)$.
 
In the phase space framework which
we are aiming at in this paper both the
$(q,p)$ variables as well as the spin variables ($\a,\bar\a$)
of the original internal spin space $\C_2$ and its complex
conjugate are regarded as
{\it phase space variables} transforming all, except for the
translations affecting only $q$, in a similar manner under Poincar\'e
transformations $(b,\Lambda)$. This property will be used in Sect.~II
to define a one-particle resolution kernel Hilbert space, 
$\H^{(s)}_{\tilde\eta}$,
for free particles of mass $m$ and arbitrary integer or half integer
spin $s$ possessing
physically reasonable relativistic localization properties, which
are described in terms of scalar functions, $\psi^{(s)}(q,p,\a)$,
obeying Eqs.~(\ref{11}), (\ref{17}) and (\ref{18}) and
transforming irreducibly
under a unitary phase space representation, $U^{(s)}(b,\Lambda)$,
of the Poincar\'e group. In Sect.~III we extend
this description to a first quantized soldered Hilbert bundle $\Hms$
over a curved Riemann-Cartan space-time base $U_4$ in order to
include and investigate influences due to gravity.
The bundle $\Hms$ is associated to
the affine spin frame bundle $P(U_4,\bar G=\bar\P)$. In
Sect.~IV we then study the quantum propagation on $\Hms$ and
define a generalized path integral formula for particles with spin.
Finally, Sect. V is devoted to some concluding remarks.
 
\vspace{1cm}

\section{One-Particle Stochastic Phase Space Description
Including Spin}
 
In this section we develop the stochastic phase space representation
of $\P$ for particles of arbitrary spin by using the internal spinor
variables $\alpha$ and $\bar\a$
introduced in the introduction. To define the notation, we begin by
briefly reviewing the spin zero case treated in detail in \cite{prug1},
and then investigate an extended framework for the stochastic phase space
description of one-particle states possessing arbitrary but
unspecified spin. We then reduce this representation into irreducible
components to yield a description for free relativistic particles
of a definite physical spin, $s=0,\ha,1,\frac{3}{2},\ldots$,
and a fixed mass value $m$.\\
 
\noindent{\bf A) The spin-zero case}
 
The aim is to construct a unitary irreducible phase space representation
of the Poincar\'e group in terms of generalized
quantum mechanical wave functions, $\psi (q,p)$, which represent a
relativistic spin-zero particle (or antiparticle) with stochastic localization
properties in the configuration space variable $q=(q^\mu;\mu=0,1,2,3)$
as well as in the momentum space variable $p=(p^\mu;\mu=0,1,2,3)$,
with $p_\mu p^\mu=m^2c^2$, where $m$ is the mass of the particle.
The reason for introducing the stochastic phase space variables
$(q,p)$ for the description of particles in high energy physics
is the impossibility to localize relativistic particles in terms
of PV-measures on Borel sets over configuration space alone, with
the operators transforming under a unitary irreducible representation
of $\P$ acting in the respective Hilbert space of states carrying this
system of imprimitivity. It is, however, possible to construct a
{\it generalized system of imprimitivity}
(called a system of covariance) in terms of POV-measures
on Borel sets over Minkowski space {\it and} momentum space, realized on a
Hilbert space $\H_{\tilde\eta}$ of states transforming under a
stochastic phase space representation $U(b,\Lambda)$ of $\P$.
This is achieved by defining irreducibly transforming one-particle
states over {\it relativistic stochastic phase space}, constructed
in terms of a resolution generator $\tilde\eta =\tilde\eta_l$
for the particle in question, with $\tilde\eta_l$ being parametrized by
an elementary length parameter $l$. Thereby Wigner's 1932
phase space formulation of quantum mechanics\cite{EPW} is turned
into (i) a fully relativistic formulation, and (ii) a formalism
possessing a probability interpretation for the description based on the
stochastic variables $q$ and $p$. The outcome of this endeavour
is the construction of a resolution kernel Hilbert space,
$\H_{\te}$, carrying a unitary irreducible spin-zero phase
space representation, $U(b,\Lambda)$, of the Poincar\'e
group, and containing states ${\bf \Psi}$
with physically reasonable (stochastic)
localization properties in the variables $q$ and $p$. For a detailed
description of this whole approach we refer to the extended work of
E.~Prugove\v{c}ki (see Refs.~\cite{prug1,prug2} and the references
quoted there) as well as to the review paper by S.T.~Ali \cite{Ali2}.
 
To define the notation we introduce the relativistic one-particle phase
space
\begin{equation}
\M^\pm_m = M_4 \times V^\pm_m\/,
\label{21}
\end{equation}
where $M_4=R_{1,3}$ denotes Minkowski space-time having the metric tensor
$\eta_{\mn}={\rm diag} (1,-1,-1,-1)$, and $V^\pm_m$ is the positive
energy $(V_m^+$ with $p^0>0)$ or negative energy ($V^-_m$ with
$p^0<0$) hyperboloid in momentum space, $p_\mu p^\mu=m^2c^2$,
possessing the Lorentz invariant measure
\begin{equation}
d\Omega_m(p) =\frac{d^3 p}{2p^0}\/~.
\label{22}
\end{equation}
The momentum space wave function for a spin-zero particle of mass
$m$ will be donoted by $\hat\psi (k)$,
with $\hat\psi (k)\in L^2 (V^\pm_m), k\in V^\pm_m$,
being square integrable with respect to the measure
(\ref{22}). The space $L^2(V^\pm_m)$ carries a unitary irreducible
representation $\hat U(b,\Lambda)$ of the Poincar\'e group:
\begin{equation}
(\hat U(b,\Lambda)\hat\psi)(k) =e^{\frac{i}{\hbar}b\cdot k}
\hat\psi (\Lambda^{-1}k)
\label{23}
\end{equation}
leaving the scalar product
\begin{equation}
\langle\hat\psi_1\mid\hat\psi_2\rangle_{V^\pm_m}=
\int_{V^\pm_m}\hat\psi_1^*(k)\hat\psi_2(k)d\Omega_m (k)
\label{24}
\end{equation}
invariant. The notation used implies that both states with momentum
wave functions $\hat\psi_1(k)$ and $\hat\psi_2 (k)$ refer either
to a particle (integration over $V^+_m$) or to an antiparticle
(integration over $V^-_m$).
 
The next step is the construction of an isometric
map, called $\W_{\te}$, between the Hilbert space $L^2(V^\pm_m)$
and a Hilbert space $L^2(\Sigma^\pm_m)$ defined over relativistic
phase space, where $\Sigma^\pm_m = \sigma \times V^\pm_m\subset\M^\pm_m$,
with $\sigma$ being a space-like hypersurface in Minkowski space,
and with the Poincar\'e invariant measure on
$\Sigma^\pm_m$ being given by
\begin{equation}
d\Sigma_m(q,p)=2\ve(p^0)p^\mu d\sigma_\mu\delta (p^2-m^2c^2)d^4p~.
\label{25}
\end{equation}
This is achieved with the help of the map
\begin{equation}
\W_{\te}: \quad  L^2(V^\pm_m)\to L^2(\Sigma^\pm_m)
\label{26}\end{equation}
defined by the following integral transform:
\begin{equation}
\psi(q,p)=(\W_{\te}\hat\psi)(q,p)=\int\limits_{V^\pm_m}
\te^*_{q,p}(k)\hat\psi (k)d\Omega_m(k)\/.
\label{27}
\end{equation}
where the $\te_{q,p}(k)$ denote a set of coherent states,
obtained from
the resolution generators $\te(k)\in L^2(V^\pm_m)$
with the help of
the Poincar\'e transformation $\hat U(q,\Lambda_p)$, involving a
translation by $q$ and a Lorentz boost with $v=p/m$ denoted by
$\Lambda_p$. Using (\ref{23}) one has
\begin{equation}
\te_{q,p}(k)=(\hat U(q,\Lambda_p)\te)(k)=e^{\frac{i}{\hbar}q\cdot k}
\te (\Lambda^{-1}_p k)\/.
\label{28}\end{equation}
Here $\te (k)$ is the resolution generator, being $SO(3)$ invariant,
i.e.~obeying, with $R\in SO(3)$ and
$\Lambda (R)=\left(\begin{array}{ll}
       \displaystyle1&0 \\
       \displaystyle0&R \end{array}\right)$,
\begin{equation}
\te (\Lambda (R) k) = \te (k)\/.
\label{29}\end{equation}
It is easy to show that (\ref{29}) implies that $\te(\Lambda^{-1}_p k)=
\eta (p\cdot k)$ with real $\eta =\eta_l$ (compare Ref.~\cite{prug1},
Chapter 2; we suppress the suffix $l$ in the sequel).
 
Eq.~(\ref{28}) defines a set of generalized coherent states in
$L^2(V^\pm_m)$ parametrized in terms of the coset $\P/SO(3)$.
So we may, finally, write the integral transform $\W_{\te}$
introduced in (\ref{27}) as
\begin{equation}
\psi(q,p)=(\W_{\te}\hat\psi)(q,p)=\int\limits_{V^\pm_m}
          e^{-\frac{i}{\hbar}q\cdot k}\eta^*(p\cdot k)
          \hat\psi (k)d\Omega_m (k)
\label{210}\end{equation}
where the complex conjugation symbol on $\eta$ is, actually,
unnecessary but we keep it for the later generalization of
Eq.~(\ref{210}). By construction the right-hand side
of (\ref{210}) satisfies the Klein-Gordon equation
in the variable $q$.
 
Mapping the coherent states (\ref{28}) into $L^2(\Sigma^\pm_m)$,
i.e.\ defining
\begin{equation}
\phi_{q,p}(q',p')=(\W_{\te}\te_{q ,p})(q',p')=\int\limits_{V^\pm_m}
   \te_{q',p'}^* (k)\te_{q,p}(k)d\Omega_m(k)\label{211}\\
   =\langle\te_{q',p'}\mid\te_{q,p}\rangle_{V^\pm_m}\/,\nonumber\\
\end{equation}
yields the result, because of the isometry property of the map
$\W_{\te}$, that the overlap between the coherent states computed in
$L^2(V^\pm_m)$ may be expressed as
\begin{equation}
\phi_{q,p}(q',p')=\langle\te_{q',p'}\mid\te_{q,p}\rangle_{V^\pm_m}
=\langle\phi_{q',p'}\mid\phi_{q,p}\rangle_{\Sigma^\pm_m}\/,
\label{212}\end{equation}
which is identical to the propagator in $L^2(\Sigma^\pm_m)$ given by
\begin{eqnarray}
K_{\te}(q',p';q,p)&=&\langle\phi_{q',p'}
\mid\phi_{q,p}\rangle_{\Sigma^\pm_m}~\nonumber\\
&=&\int\limits_{\Sigma^\pm_m}\phi^*_{q',p'}(q'',p'')
    \phi_{q,p}(q'',p'')d\Sigma_m(q'',p'')\/,
\label{213}
\end{eqnarray}
where the second equality defines the scalar product in $L^2(\Sigma^\pm_m)$.
Using the fact that the states $\phi_{q,p}$
allow the following resolution of the
identity in $L^2(\Sigma^\pm_m)$
\begin{equation}
\int\limits_{\Sigma^\pm_m}\vert\phi_{q,p}\rangle d\Sigma_m(q,p)\langle
\phi_{q,p}\vert={\bf 1}^\pm\/,
\label{214}\end{equation}
we see that
$K_{\te}(q',p';q,p)$ obeys the following
reproducing and reality relations implied by
(\ref{213}):
\begin{equation}
K_{\te}(q',p';q,p)=\int\limits_{\Sigma^\pm_m}K_{\te}
(q',p';q'',p'')K_{\te}(q'',p'';q,p)d\Sigma_m(q'',p'')~,
\label{215}\end{equation}
and
\begin{equation}
K^*_{\te}(q',p';q,p)=K_{\te}(q,p;q',p')=\phi^*_{q',p'}(q,p)\/.
\label{216}\end{equation}
 
Any state ${\bf \Psi}\in\H_{\te}\equiv L^2(\Sigma^\pm_m)$ may now
be decomposed in terms of the states $\phi_{q,p}$ providing
a coherent state basis in $\H_{\te}$. The result is
\begin{equation}
{\bf \Psi}=\int\limits_{\Sigma^\pm_m}\psi(q,p)\phi_{q,p}d\Sigma_m(q,p)\/,
\label{217}\end{equation}
where $\psi (q,p)=\langle\phi_{q,p}\vert {\bf \Psi}\rangle_{\Sigma^\pm_m}$
is a generalized one-particle relativistic
quantum mechanical wave function (a scalar field on stochastic
phase space) transforming under Poincar\'e transformations $(b,\Lambda)$
in the following manner:
\begin{equation}
(U(b,\Lambda)\psi)(q,p)=\psi(\Lambda^{-1}(q-b),\Lambda^{-1}p)\/.
\label{218}\end{equation}
Eq.~(\ref{218}) is easily proven by applying  (\ref{23}) in
(\ref{210}) and making use of the invariance of the measure
(\ref{22}). Thus $\W_{\te}$ is an intertwining operator for
the representations $\hat U(b,\Lambda)$ and $U(b,\Lambda)$ obeying
\begin{equation}
U(b,\Lambda)\W_{\te}=\W_{\te}\hat U(b,\Lambda).
\label{219}\end{equation}
 
Using (\ref{213}) and (\ref{214}) in the definition of $\psi(q,p)$
given above one easily derives the following propagation formula:
\begin{equation}
\psi(q',p')=\int\limits_{\Sigma^\pm_m} K_{\te}(q',p';q,p)\psi(q,p)
d\Sigma_m(q,p)~.
\label{2190}\end{equation}

The phase space representation $U(b,\Lambda)$ of $\P$
defined by (\ref{210}) and (\ref{218}) leaves invariant the following scalar
product in $\H_{\te}$ obtained from (\ref{213}) and (\ref{217}):
\begin{equation}
\langle\psi_1\mid\psi_2\rangle_{\Sigma^\pm_m}=
\int\limits_{\Sigma^\pm_m=\sigma\times V^\pm_m}\psi^*_1(q,p)
\psi_2(q,p)d\Sigma_m(q,p)\/.
\label{220}\end{equation}
Eq.~(\ref{220}) may also be written as
\begin{equation}
\langle\psi_1\mid\psi_2\rangle_{\Sigma^\pm_m}=\frac{i\hbar}{Z_{\te}}\int\int
      \psi^*_1(q,p)\buildrel\leftrightarrow \over \pa_\mu \psi_2(q,p)
      d\sigma^\mu (q) d\Omega_m (p)
\label{221}\end{equation}
with $\pa_\mu =
\pa/\pa q^\mu$, and with
\begin{equation}
Z_{\te} =(2\pi\hbar)^3\int\limits_{V^\pm_m}\vert\eta (p\cdot k)\vert^2
      d\Omega_m (p)
\label{222}\end{equation}
being a constant independent of $k$. For a particular choice
of the resolution generator $\eta$ this
yields an irreducible unitary representation
$U(b,\Lambda)$ on $\H_{\te}$ describing spin-zero
particles of mass $m$ (see Refs.~\cite{prug1} and \cite{prug2}).
In our context it is essential to observe that the resolution
generator introduces a particular smearing in the variables $q$ and
$p$ (in accordance with Heisenberg's uncertainty relations)
which is parametrized here in terms of a fundamental length paramter
$l$ for the particular type of particles involved with
$\lbrack m,s\rbrack =\lbrack m,0\rbrack$.
The actual value of $l$, i.e.\ whether it is of order of
$10^{-16}$~cm, i.e. well below the charge radius of a nucleon,
or even equal to the Planck length $\sim 10^{-32}$~cm
is not essential in the present context. The main point is the
regularizing effect this length parameter has in the stochastic phase
space formalism. Taking, however,
the sharp point limit $l\to 0$ leads to the
appearance of singularities in Eqs.~(\ref{221}) and (\ref{222})
(compare Refs.~\cite{prug1} and \cite{prug2}).
This is reminiscent of the situation prevailing in the
conventional relativistic quantum field theory
based on $q$-space fields obtained by ordinary Fourier transformation
from the $p$-space fields.
The stochastic phase space description introduced
in Refs.~\cite{prug1}\cite{prug2} and \cite{Ali2} was just proposed in
order to avoid the singularities of the conventional formalism.
We shall thus assume the fundamental length paramter $l$ to have
a small but finite fixed value.
 
Our task now is to extend the spin-zero formalism reviewed above
to the case of a particle with arbitrary spin $s$. This will be done
in the following subsection by using the internal spin space variables
$\alpha$ and $\bar\alpha$ for a homogeneous space description
of spin as described in the introduction.
\\
 
\noindent{\bf B) The non-zero spin case}
 
In view of the discussion presented in the introduction we
represent a relativistic particle with arbitrary but unspecified
spin and definite mass $m$ by a scalar wave function $\hat\psi (k,\alpha,
\bar\alpha)$ defined on momentum and spin space, $V^\pm_m\times\C_2$,
with the invariant measure (\ref{22}) on $V^\pm_m$ and the
invariant measure $d\alpha d\bar\alpha$ on $\C_2$ \cite{texta}
(compare the remarks made after (\ref{16}) above).
 
As regards the transformation rule for the spinor variables
$\alpha =(\alpha^1,\alpha^2)$ and $\bar\alpha=(\bar\alpha^{\dot 1},
\bar\alpha^{\dot 2})$ introduced in Sect.~I we observe that, conventionally,
a spinor with a lower undotted spinor index $A$ is taken to transform
with the $SL(2,\C)$ matrix $D^{(\ha,0)}(\Lambda)$, while an upper
dotted spinor index $\dot A$ transforms with the matrix
$\lbrack D^{(\ha,0)}(\Lambda^{-1})\rbrack^\dagger=D^{(0,\ha)}(\Lambda)$
(compare, for example, Carruthers \cite{Carr}). This leads for our
spinor variables $\alpha,\bar\alpha$ with only upper indices to
the following transformation rules (written as matrix operation from the
left and with $T$ denoting the transpose):
\begin{eqnarray}
\alpha'=D(\Lambda)\alpha\quad\mbox{with}\quad
      D(\Lambda)&=&\lbrack D^{(\ha,0)}(\Lambda^{-1})\rbrack^T
\label{223a}\\
\bar\alpha'=\bar D(\Lambda)\bar\alpha \quad\mbox{with}\quad
\bar D(\Lambda)&=&D^{(0,\ha)}(\Lambda) \label{223b}
\end{eqnarray}
 
The generalization of Eq.~(\ref{23}) in the presence of spin
described through the internal spinor variables $\alpha,\bar\alpha$
now reads
\begin{equation}
(\hat U(b,\Lambda)\hat\psi)(k,\alpha,\bar\alpha)=
e^{\frac{i}{\hbar} b\cdot k} \hat\psi (\Lambda^{-1}k,D(\Lambda^{-1})
\alpha,\bar D(\Lambda^{-1})\bar\alpha)
\label{224}\end{equation}
where, of course, $\hat U(b,\Lambda)$
is not an irreducible representation
here. Before we discuss the reduction of the function $\hat\psi(k,\alpha,
\bar\alpha)$ to one describing a particle with definite spin value $s$
we also generalize the scalar product defined in (\ref{24}) to functions
defined on $V^\pm_m\times \C_2$ \cite{texta}:
\begin{equation}
\langle\hat\psi_1\mid\hat\psi_2\rangle_{V^\pm_m\times\C_2} =
\int\limits_{V^\pm_m\times\C_2} \hat\psi_1^*(k,\alpha,\bar\alpha)
\hat\psi_2(k,\alpha,\bar\alpha)d\Omega_m(k)d\alpha d\bar\alpha~.
\label{225}\end{equation}
Due to (\ref{224}) and the
invariance of the measure on $V^\pm_m\times\C_2$
this is a Poincar\'e invariant scalar product if it exists.
We shall assume here that the momentum space wave function
$\hat\psi (k,\alpha,\bar\alpha)$ of a spinning particle is,
indeed, square-integrable over $V^\pm_m$ {\it and} $\C_2$ requiring
that the high spin states contained in $\hat\psi$ (see the reduction
described below) are sufficiently damped to
compensate for the exponential
factor $e^{2t}$ in the measure (\ref{16}) when expressed in the
variables ($t,\psi,\theta,\varphi$). In the conventional formulation
(compare Ref.~\cite{AliP}) a decomposition of the
original Hilbert space $\hat\H$ into an infinite direct sum of
irreducible subspaces $\hat\H^{(s)}=L^2(V^\pm_m)\otimes K_s$, with $K_s$
being a spin space of dimension $2s+1$, is considered, i.e.
\begin{equation}
\hat\H=\sum^\infty_{s=0}\oplus\hat\H^{(s)}~,
\label{226}\end{equation}
where arbitrary large spin values $s$ are involved. In the
present context we shall identify $\hat\H$ with the Hilbert space
$L^2(V^\pm_m\times\C_2)$ defined by (\ref{224}) and (\ref{225})
assuming $\hat\psi (k,\alpha,\bar\alpha)$ to be square-integrable
with respect to the measure $d\Omega_m(k)d\alpha d\bar\alpha$.
 
Before we go on to construct the analogue of the map $\W_{\te}$
in the present case yielding an irreducible subspace $\H^{(s)}_{\tilde\eta}$
of the Hilbert space $L^2(\Sigma^\pm_m\times\C_2)$ introduced below,
let us reduce the general $\C_2$-description
of spin by demanding homogeneity and holomorphicity in the spinor
variables as expressed by Eqs.~(\ref{17}) and (\ref{18}).
As described in the introduction, this amounts to the following
restrictions:
\begin{eqnarray}
\hat\psi (k,\alpha,\bar\alpha)&\tohoch^{(1.7),(1.8)}&
\hat\psi^{(s)}(k,\alpha)\/,\nonumber\\
\C_2 &\tohoch^{(1.9)}&S^2_{r=2s}\/.\label{227}
\end{eqnarray}
The reduction to a definite spin value will thus be
governed by the equations
\begin{eqnarray}
\hat D\hat\psi (k,\alpha,\bar\alpha) &=&2s
    \hat\psi (k, \alpha,\bar\alpha)~, \label{228}\\
    \frac{\pa}{\pa\bar\alpha^{\dot A}} \hat\psi (k,\alpha,\bar\alpha)
   &=&0\,;\quad\dot A=\dot 1,\dot 2~, \label{229}
~~{\rm and}
\\
k_{\dot A\dot A}\alpha^A\bar\alpha^{\dot A}&=&mcr\,;\quad\mbox{with $r$
being an integer.}\label{230}
\end{eqnarray}
A scalar momentum space wave function satisfying (\ref{228})-(\ref{230})
will be denoted, according to (\ref{227}), by $\hat\psi^{(s)}(k,\alpha)$.
Here $\hat D=\alpha^A\pa/\pa\alpha^A$ represents an
invariant operator which commutes with $\hat U(b,\Lambda)$ as
defined in (\ref{224}); the same is true for the invariant
operator appearing on the l.-h.\ side of
(\ref{230}). Using Eqs.~(\ref{15}) and their complex conjugates,
Eq. (\ref{230}) may be written as
\begin{equation}
e^t(k^o + \vec n\cdot \vec k)=mcr\/,
\label{231}\end{equation}
where $\vec n$ is a unit vector given by
\begin{equation}
\vec n = (\sin\theta\cos\varphi,\sin\theta\sin\varphi,\cos\theta)~.
\label{232}\end{equation}
In the derivation of (\ref{231}) the angle $\psi$  has disappeared
in accordance with the remarks concerning the $U(1)$ degree of freedom
made around Eq.~(\ref{19}) in the introduction.
In the rest system of the particle we thus have in the
reduced case from (\ref{231}) that
\begin{equation}
e^t = r = 2s =\mbox{ fixed integer }\/.
\label{233}\end{equation}
This constrains the integration over the $\C_2$ variables,
for example, in (\ref{225}) and in analogous equations derived below
in the reduced case where Eqs.~(\ref{228})-(\ref{230}) are to be
satisfied.
Let us now decompose $\hat\psi^{(s)}(k,\alpha)$ in view of
(\ref{228}) into a homogeneous polynomial of degree $2s$
in $\alpha^1$ and $\alpha^2$
yielding (compare Eqs.~(\ref{15}))
\begin{equation}
\hat\psi^{(s)}(k,\alpha)=\sum_{s_3}\hat\psi^{(s)}_{s_3} (k)
~r^s\frac{(\ual^1)^{s+s_3}(\ual^2)^{s-s_3}}
{\sqrt{(s+s_3)!(s-s_3)!}}
\label{234}\end{equation}
where the sum over $s_3$ runs from $s_3=-s$ to $s_3=+s$.
Moreover, we have written the $t$-dependence explicitly as
$(e^t)^s=r^s$ and denoted the angular part depending on
$\psi,\theta$, and $\varphi$ by underlined spinor components.
Following Edmonds \cite{Ed} we call the appearing normalized
homogeneous polynomials $\uss$, i.e.
\begin{equation}
\uss=\frac{{(\ual^1)}^{s+s_3}{(\ual^2)}^{s-s_3}}
         {\sqrt{(s+s_3)!(s-s_3)!}}~,
\label{235}\end{equation}
possessing the following well-known behaviour under
$SO(3)$-rotations $R_{\abg}$ (compare also
Ref. \cite{Wigner} p. 163 - 165)
\begin{equation}
U(R_{\abg})\uss=u'(s,s_3)=\sum_{s_3'}\usp D^s_{s'_3,s_3}(\abg)
\label{236}\end{equation}
where $u'(s,s_3)$ is the expression (\ref{235}) computed with the
rotated spinor components $\ual'^A$; $A=1,2$, and
$D^s_{s'_3,s_3}(\abg)$ are the Wigner $D^s$-functions.
 
We now consider Eq.~(\ref{224}) for the reduced function
$\hat\psi^{(s)}(k,\alpha)$. Using the
decomposition (\ref{234}) yields
\begin{equation}
\sum_{s_3} (\hat U(b,\Lambda)\hpss)(k)
~r^s\uss=\sum_{s_3}e^{\frac{i}{\hbar}b\cdot k}\hat\psi^{(s)}_{s_3}
(\Lambda^{-1}k)~ r^s u'(s,s_3)
\label{237}\end{equation}
where here $u'(s,s_3)$ is constructed, according to (\ref{223a})
and (\ref{233}), in terms of the spinor
\begin{equation}
\ual'=D(\Lambda^{-1})\ual\,,\mbox{ with }\ual =
e^{-\frac{t}{2}}\alpha,\ \ual'=e^{-\frac{t}{2}}\alpha'\/.
\label{238}\end{equation}
Thus $u'(s,s_3)$ differs from $\uss$ by a rotation
associated with the transition from $k$ to $\Lambda^{-1}k$.
This is the Wigner rotation (compare (\ref{110}))
\begin{equation}
R^{-1}(k,\Lambda^{-1})=\Lk\Lambda\Lambda_{\Lambda^{-1}k}\/.
\label{239}\end{equation}
Using this rotation in (\ref{236}) and inserting the
resulting expression for $u'(s,s_3)$ into the r.-h.\ side
of (\ref{237}) we conclude -- taking, moreover, the orthonormality
of the $\uss$ into account -- that the following transformation rule
for $\hpss (k)$ must hold:
\begin{equation}
\Bigl(\hat U^{(s)}(b,\Lambda)\hpss\Bigr)(k)=
e^{\frac{i}{\hbar}b\cdot k}\sum_{s_3'}
D^s_{s_3,s'_3}(\Lk\Lambda\Lambda_{\Lambda^{-1}k})
\hat\psi^{(s)}_{s'_3}(\Lambda^{-1}k)~.
\label{240}\end{equation}
This is the typical transformation law for the momentum eigenstate
of a free particle of spin $s$ and spin projection $s_3$,
derived here from Eq.~(\ref{224}) for a definite integer or half integer
spin value $s$. Moreover, in (\ref{240}) we have denoted by
$\hat U^{(s)}(b,\Lambda)$ the irreducible action of $\hat U(a,\Lambda)$
in the $(2s+1)$-dimensional vector space defined by $\hat \psi^{(s)}_{s_3}(k)$.
 
Considering now the square of the
wave function $\hat\psi^{(s)}(k,\alpha)$
according to (\ref{225}), and taking recognition of the constraint
(\ref{230}) by introducing a $\delta$-function
\begin{equation}
\delta(\frac{1}{mc}k_{A\dot A}\alpha^A\bar\alpha^{\dot A}-r),\quad
\mbox{with integer } r= 2s
\label{241}\end{equation}
into the $\C_2$-integration, yields with the help of
(\ref{233})
\begin{equation}
\langle\hps\mid\hps\rangle_{\Vpmm}=
\int\limits_{\Vpmm}\lbrack \hps (k,\a)\rbrack^*\hps (k,\a)\delta
\Bigl(\frac{1}{mc}k_{A\dot A}\a^A\bar\a^{\dot A}-r\Bigr)d\Omega_m(k)
d\a d\bar\a\/.
\label{242}\end{equation}
This may be written in terms of the coordinates $(t,\psi,\theta,\varphi)$
in using (\ref{15}), (\ref{16}) together with (\ref{234}) as
\begin{eqnarray}
&\langle&\hps\mid\hps\rangle_{\Vpmm}\nonumber\\
&=&\int\limits_{\Vpm} d\Omega_m(k)
\int\sum_{s_3s_3'}\lbrack\hps_{s'_3}(k)\rbrack^*\hpss (k)
r^{2s}\lbrack\usp\rbrack^*\uss\delta(e^t-r)e^{2t}dtd\psi\sin\theta d\theta
d\varphi \/.
\label{243}
\end{eqnarray}
Here $\uss$, defined in (\ref{235}), is expressed in terms of
$\ual^A(\psi,\theta,\varphi);A=1,2$. We can use Eq.~(\ref{236})
to write $\uss$  in terms of $D^s$-functions, i.e.
\begin{equation}
\uss =\sum_{s_3'}\buildrel\circ\over{u}\!(s,s'_3) D^s_{s'_3s_3}(\psi,
\theta,\varphi)
\label{244}\end{equation}
where $\buildrel \circ\over{u}\!(s,s_3)$ is the homogeneous
polynomial constructed with
$\ual^1=1$ and $\ual^2=0$ being different from zero only for $s=s_3$.
With the help of the familiar
result\cite{Ed}
\begin{equation}
\frac{1}{8\pi^2}\int\limits^{2\pi}_0\int\limits^\pi_0\int\limits^{2\pi}_0
D^{j_1*}_{m'_1m_1}(\psi,\theta,\varphi)
D^{j_2}_{m_2'm_2}(\psi,\theta,\varphi)
d\psi\sin\theta d\theta d\varphi =   \label{246}
\delta_{m'_1m'_2}\delta_{m_1m_2} \delta_{j_1j_2}\frac{1}{2j_1+1}
\nonumber
\end{equation}
as well as the normalization of the $\buildrel\circ\over{u}\!(s,s_3)$,
one finally obtains
\begin{equation}
\langle\hps\mid\hps\rangle_{\Vpmm}=\frac{8\pi^2}{2s+1}r^{2s+1}
\int\limits_{\Vpm}\sum_{s_3}\vert\hpss (k)\vert^2 d\Omega_m (k)\/.
\label{247}\end{equation}
Here we could now absorb the constants appearing
in front of the integral into
the normalization of the wave functions $\hpss (k)$; $s_3=-s\ldots +\!s$,
for each particular spin value $s=\ha,1,\frac{3}{2},2\ldots$.
Summing up we may say that
the reduction to a definite spin value $s$ governed by
Eqs. (\ref{227})-(\ref{230}) thus leads to wave functions $\hpss (k)$
being elements of the Hilbert space 
$\hat\H^{(s)}=L^2(V^{\pm}_m)\otimes K_s$. As regards spin they
transform irreducibly under the representation of $SL(2,\C)$
usually denoted by $D^{(s,0)}$.

In defining now, in the presence of spin, a map $\tilde\W_{\te}$ from
$L^2(\Vpmm)$ to $L^2(\Sigma^\pm_m\times\C_2)$,
and constructing a unitary reducible phase space representation for
particles with arbitrary spin in terms of scalar functions on generalized
phase space, we make the following observation concerning the
variables $\alpha$ and $\bar\alpha$. In fact, the pair $(\a,\bar\a)$
may be regarded as phase space variables for spin in analogy to
$(q,p)$ being the phase space variables for the kinematic localization
of spin-zero particles. Quantum mechanically the momentum operators
(for the phase space representation) are constructed with the operators
$i\hbar\pa/\pa q^\mu$
producing the eigenvalue $p_\mu$, while the spin operator
(or ``spin measuring operator'') $\hbar\hat D$, producing the
eigenvalue $2s\hbar$, is constructed in terms of
$\hbar\pa/\pa\a^A$.
So one could regard $\alpha^A$ as a position-type variable for spin
and $\bar\a^{\dot A}$ as the corresponding conjugate momentum-type
variable for spin. (Of course, we already know from the discussion
presented above that in a quantized theory describing free particles
of definite spin the $\bar\alpha$ variables disappear and only
half the spin variables remain to describe a particle of definite
spin.) Hence, in defining the integral transform $\tilde\W_{\te}$
in the presence
of spin (and prior to the reduction), an invariant integration over
momentum-type variables must be involved -- with spin included!
One would therefore expect -- provided the
mentioned analogy between ordinary
phase space variables and spin variables is indeed correct -- that
an integration over $\bar\a$ is involved
in generalizing Eq.~(\ref{27}) to the non-zero spin case. We thus
propose the following integral transform to yield
$\psi(p,q;\alpha,\bar\a)$:
\begin{eqnarray}
\psi(q,p;\a,\bar\a)&=&(\tilde\W_{\te}\hat\psi)(q,p;\a,\bar\a)\label{248}\\
&=&\frac{1}{N}\int\limits_{\Vpm}\int\limits_{\bar\a'}\lbrack\te_{q,p}
   (k,\a,\bar\a')\rbrack^*\hat\psi (k,\a,\bar\a')d\Omega_m(k)
   d\bar\a'\/,\nonumber
\end{eqnarray}
where the coherent state basis for non-zero spin is, in analogy to
(\ref{28}) and in view of (\ref{224}),
\begin{eqnarray}
\te_{q,p}(k,\a,\bar\a)&=&(\hat U(q,\Lambda_p)\te)(k,\a,\bar\a) \label{249}\\
&=&e^{\frac{i}{\hbar}q\cdot k}\te(\Lp k,D(\Lp)\a, \bar D(\Lp)\bar\a)
\/.\nonumber
\end{eqnarray}
The measure $d\bar\a'$ in (\ref{248}) is $\bar\P$-invariant
due to the unimodularity of the group $SL(2,\C)$. $N$ in (\ref{248})
is a normalization constant associated with the integration over $\bar\a'$.
 
It is now essential to remark that
the resolution generator $\te (k,\a,\bar\a)$ in (\ref{249})
is again assumed to
be $SO(3)$-invariant, i.e.\ generalizing Eq.~(\ref{29})
the following relation holds
\begin{equation}
\te\Bigl(\Lambda (R)k, D(R)\a, \bar D(R)\bar\a\Bigr)=\te (k,\a,\bar\a)\/,
\label{250}\end{equation}
where $D(R)$ denotes the $SL(2,\C)$ matrix corresponding to
a rotation $R\in SO(3)$, with $D(R)$ and $\bar D(R)$ denoting
thus {\it equivalent\/} representations of $SU(2)$ as is well-known.
 
It is easy to show using (\ref{224}) and the rotation invariance
(\ref{250}) of $\te$ that the intertwining relation
(\ref{219}) for $\tilde\W_{\tilde\eta}$
is again valid with $U(b,\Lambda)$ acting on the states
$\psi (q,p;\a,\bar\a)$ in the following way:
\begin{equation}
(U(b,\Lambda)\psi)(q,p;\a,\bar\a)=\psi(\Lambda^{-1}(q-b),
\Lambda^{-1} p; D(\Lambda^{-1})\a,\bar D(\Lambda^{-1})\bar\a)\/.
\label{251}\end{equation}
In order to establish (\ref{251}), using (\ref{248}),
the spinor $D(\Lp)\a$ appearing
in the argument of $\te$ in $\tilde\W_{\te}\hat U(b,\Lambda)\hat\psi$
is written as
\begin{equation}
D(\Lp)\a = D(R^{-1}(p,\Lambda^{-1})) D(\Lambda^{-1}_{\Lambda^{-1}p})
D(\Lambda^{-1})\alpha
\label{252}\end{equation}
with $R^{-1}(p,\Lambda^{-1})$ as given by (\ref{239}). In the argument
of $\te$ the Wigner rotation may however
be dropped due to the $SO(3)$-invariance
(\ref{250}). This has the consequence that the spinor variables of
$\te$ ``feel only the boosts'', acting in {\it inequivalent\/}
ways on $\alpha$ and $\bar\a$, establishing thus, finally,
Eq.~(\ref{251}).
 
A Poincar\'e invariant scalar product for the phase space wave
functions $\psi (q,p;\alpha,\bar\a)$ satisfying (\ref{251})
may now be written down generalizing Eq.~(\ref{220}):
\begin{equation}
\langle\psi_1\mid\psi_2\rangle_{\Sigma^\pm_m\times\C_2}=
\int\limits_{\Sigma^\pm_m\times\C_2}\lbrack\psi_1
(q,p;\a,\bar\a)\rbrack^*\psi_2(q,p;\a,\bar\a)d\Sigma_m(q,p)
d\a d\bar\a
\label{253}\end{equation}
 
Let us next investigate Eqs.~(\ref{248}) and (\ref{249}) in the
reduced case assuming Eqs.~(\ref{227})-(\ref{230}) to hold true.
In this case the $\bar\a'$ dependence of $\te$ and $\hat\psi$
disappears and the integration over $d\bar\a'=e^{\ha t}d\bar{\ual}'$
represents, in view of (\ref{233}), an angular integration which can be
carried out. Adjusting the constant $N$ appropriately this yields
\begin{eqnarray}
\psi^{(s)}(q,p;\a,\bar\a)&=&\left(\tilde\W_{\te}\hat\psi^{(s)}\right)
  (q,p;\a,\bar\a)\nonumber\\
&=&\int\limits_{\Vpm}e^{-\frac{i}{\hbar}q\cdot k}
  \lbrack{\te}^{(s)}(\Lp k,D(\Lp)\a)\rbrack^*
  \hat\psi^{(s)} (k,\a)d\Omega_m (k)~,\label{254}
\end{eqnarray}
where $\hat\psi^{(s)}(k,\alpha)$ is the homogeneous polynomial in
$\alpha$ given in (\ref{234}). Let us immediately remark
that $\psi^{(s)}(q,p;\alpha,\bar\a)$ defined by (\ref{254}) does seem
to develop now an $\bar\alpha$-dependence through the complex conjugation
of the expression in the square brackets under the integral provided the
resolution generator $\te^{(s)}$ does, indeed, depend on the spinor
variables $\alpha$. This is, however, not the case, and the r.-h.\ side
of (\ref{254}) will, in fact, define a quantity $\psi^{(s)}(q,p;\a)$
independent of $\bar\a$.
 
To see that $\te^{(s)}(\Lp k, D(\Lp)\a)$ cannot depend on $\alpha$
we note that the $SL(2,\C)$ matrix for a rotation free boost,
\begin{equation}
D^{(\ha, 0)}(\Lambda_p) =\frac{1}{\sqrt{2mc(p^0+mc)}}\lbrack mc1+p^01+
\vec\sigma\cdot\vec p\rbrack\/,
\label{255}\end{equation}
changes the real length factor $e^{\frac{t}{2}}$ of $\a$
(compare (\ref{15})). On the other hand, fixing the spin to the
value $s$ restricts this factor to $\sqrt{r}=\sqrt{2s}$
(integrality condition (\ref{233})).
This contradiction can, in view of the rotation invariance
(\ref{250}), only be avoided if $\te^{(s)}$ {\it does not depend
on $\alpha$ at all\/}. Since in this case
$\hat D\te^{(s)}=0$, it does not depend on $s$ either, and we can
replace the resolution generator appearing in (\ref{254}) by the one
describing the spin zero case in Eq.~(\ref{27}), i.e.
\begin{equation}
\te^{(s)}(\Lp k)=\eta (p\cdot k)\/.
\label{256}\end{equation}
We may thus, finally, rewrite (\ref{254}), remembering
that $\eta (p\cdot k)$ is real, as
\begin{eqnarray}
\psi^{(s)}(q,p;\a)&=&\left(\W_{\te}\hat\psi^{(s)}\right)(q,p;\a)
\nonumber\\
&=&\int\limits_{\Vpm}e^{-\frac{i}{\hbar}q\cdot k}\eta (p\cdot k)
    \hat\psi^{(s)}(k,\a)d\Omega_m (k)\/,\label{257}
\end{eqnarray}
where we have denoted the integral transform $\tilde\W_{\te}$ by the
same symbol as in the spinless case in Eq.~(\ref{210}) above.
Applying the operator $\hat D$ to both sides of this equation it is
clear that $\hat D$ commutes with the map $\W_{\te}$, i.e.
\begin{equation}
\hat D\W_{\te}=\W_{\te}\hat D\/.  \label{258}
\end{equation}
 
One can now again decompose $\hat\psi^{(s)}(k,\a)$ according to
Eq.~(\ref{234}) and define $(2s+1)$-dimensional vector-valued
phase space functions $\psi^{(s)}_{s_3}(q,p)$ in terms of
momentum space wave functions
$\hat\psi^{(s)}_{s_3}(k)$, with $(s,s_3)$; $s_3=-s\ldots +\!s$,
taking values in the spin space $K_s$ of dimension $2s+1$.
We thus see that the correspondence
(\ref{248}) yields, for the reduced states of definite spin $s$,
an isometric map (\ref{257}) -- constructed in the same manner
as in the spin-zero case -- relating the Hilbert spaces
$L^2(\Vpm)\otimes K_s\equiv\hat\H^{(s)}$
[compare (\ref{247})] and
$L^2(\Sigma^\pm_m)\otimes K_s\equiv\H^{(s)}_{\te}$. To have
a condensed notation at our disposal one can, however, express the
relations under study at first in terms of the scalar fields
$\psi^{(s)}(q,p;\alpha)$ and $\hat\psi^{(s)}(k,\alpha)$
and then go over at a later stage to the $(2s+1)$-dimensional
vector-valued fields by making an $\alpha$-expansion in terms
of homogeneous polynomials of degree $2s$.
 
Mapping the coherent state basis for the case of definite
spin $s$,
\begin{equation}
\tes (k) =\te_{q,p}(k)=e^{\frac{i}{\hbar}q\cdot k}\eta (p\cdot k)\/,
\label{259}
\end{equation}
into $L^2(\Sigma^\pm_m)$ as in Eq.~(\ref{211}) yields again
\begin{eqnarray}
\phi_{q,p}(q',p')&=&(\W_{\te}\tes) (q',p')=
    \langle\te_{q',p'}\mid\te_{q,p}\rangle_{\Vpm}\nonumber\\
&=&\langle\phi_{q',p'}\mid\phi_{q,p}\rangle_{\Sigma^\pm_m}=
    K_{\te}(q',p';q,p)\/.\label{260}
\end{eqnarray}
This implies that the stochastic phase space
propagator $K^{(s)}_{\te}(q',p';q,p)$
for a free particle of spin $s$ is the same as that for
a spin-zero particle defined in (\ref{213}). Hence, freezing the
spin content of the fields to any physical
value $s$ does not alter the phase space
kinematics of free stochastic propagation.
 
Introducing the resolution of the identity for the subspace
of definite spin $s=0,\ha, 1,\frac{3}{2}\ldots$ in $L^2(\Sigma^\pm_m\times
\C_2)$ by (\ref{214}) -- derived above for $s=0$ only but due to
(\ref{259}) and (\ref{260}) being valid generally -- one
can write down the following expansion
for a state vector $\bfPsi^{(s)}(\a)$ of definite integer or half integer
spin belonging to the Hilbert space $\H^{(s)}_{\te}$:
\begin{equation}
\bfPsi^{(s)}(\a)=\int\limits_{\Sigma^\pm_m}\psi^{(s)}(q,p;\alpha)
   \phi_{q,p}d\Sigma_m (q,p)
\label{261}\end{equation}
with
\begin{equation}
\psi^{(s)}(q,p;\alpha)=\langle \phi_{q,p}\mid\bfPsi^{(s)}(\a)
   \rangle_{\Sigma^\pm_m}\/.\label{262}
\end{equation}
The wave function $\psi^{(s)}(q,p;\alpha)$ obeys the same propagation
rule on stochastic phase space as does the spin zero wave function
$\psi (q,p)$ which is expressed by Eq. (\ref{2190}).

By construction $\psi^{(s)}(q,p;\alpha)$ is a solution of the
equations (\ref{11}), (\ref{12}), (\ref{17}) and (\ref{18})
characterized by $[m,s]$.
$\H^{(s)}_{\te}$ carries the UIR of the Poincar\'e group
denoted by $U^{(s)}(b,\Lambda)$, and
the generalized scalar one-particle phase space function
$\psi^{(s)}(q,p;\a)$ for a free
particle (or antiparticle) of mass $m$ and definite spin $s$
transforms irreducibly under Poincar\'e transformations
$U^{(s)}(b,\Lambda)$ according to (compare (\ref{251})):
\begin{equation}
(U^{(s)}(b,\Lambda)\psi^{(s)})(q,p;\alpha)=
\psi^{(s)}(\Lambda^{-1}(q-b),\Lambda^{-1}p;D(\Lambda^{-1})\alpha)
\label{263}\end{equation}
with the invariant scalar product in $\H^{(s)}_{\te}$
given by (compare (\ref{242}))
\begin{equation}
\langle\psi^{(s)}_1\mid\psi_2^{(s)}\rangle_{\Sigma^\pm_m\times\C_2}
=\int\limits_{\Sigma^\pm_m\times\C_2}
\lbrack\psi^{(s)}_1(q,p;\alpha)\rbrack^*
\psi^{(s)}_2 (q,p;\a)\delta (\frac{1}{mc}p_{A\dot A}\a^A\bar\a^{\dot A}
-r)d\Sigma_m (q,p)d\a d\bar\a\/.
\label{264}\end{equation}
 
Decomposing $\psi^{(s)}_1(q,\rho;\a)$ and $\psi^{(s)}_2(q,\rho;\a)$
into homogeneous polynomials in $\a^1$ and $\a^2$ as in Eq.~(\ref{234})
and carrying out the $\a$-integrations yields [compare (\ref{247})]
\begin{equation}
\langle\psi^{(s)}_1\mid\psi^{(s)}_2\rangle_{\Sigma^\pm_m\times\C_2}
=N_s\int\limits_{\Sigma^\pm_m}\sum\limits_{s_3}[\psi^{(s)}_{1,s_3}
(q,p)]^*\psi^{(s)}_{2,s_3}(q,p) d\Sigma_m(q,p)\label{265}
\end{equation}
with $N_s=8\pi^2~r^{2s+1}/(2s+1)$ and $r=2s$ according to (\ref{233}).
 
To conclude this section we define a system of covariance of the
Poincar\'e group (a generalized system of imprimitivity) for free
particles of definite mass $m$ and arbitrary integer or half integer
spin $s$ described by the wave function $\psi^{(s)}(q,p;\alpha)$,
defined in (\ref{257}), transforming under the irreducible unitary
phase space representation $U^{(s)}(b,\Lambda)$ of $\P$ [or rather
$\bar\P$ as far as the $\alpha$-variable is concerned;
see Eq.~(\ref{263})]
realized on the Hilbert space $\H^{(s)}_{\te}$ constructed above.
 
Due to the independence on the spin variable $s$ of Eqs.~(\ref{259}),
(\ref{260}) as well as the resolution of the identity obtained
after the reduction to a definite integer or half integer spin
[compare (\ref{214})],
one has for any $\Delta_j$ belonging to a
family of Borel sets $\B$ on {\it relativistic stochastic
phase space} \cite{AP} the following operators: A positive
operator-valued (POV) measure, $E(\Delta_j)$, on
$\H^{(s)}_{\te}$ together with a UIR of the Poincar\'e group,
$U^{(s)}(b,\Lambda)=U^{(s)}(g)$, on $\H^{(s)}_{\te}$ obeying
\begin{eqnarray}
E(\Delta_j)=E^*(\Delta_j)\geq 0;\quad&\mbox{with}\quad&E(\emptyset)=0,
\label{266}\\
E(\bigcup\limits^\infty_{j=1}\Delta_j)=\sum\limits^{\infty}_{j=1}
E(\Delta_j)\quad&\mbox{for}~&\Delta_i
\bigcap\Delta_j=\emptyset;i\not=j,
\label{267}
\end{eqnarray}
and
\begin{equation}
U^{(s)}(g)E(\Delta_j)U^{(s)}(g){}^{\dagger}=E(g\Delta_j)~,\label{268}
\end{equation}
where, for brevity, we have denoted the element $(b,\Lambda)$ of
$\P$ by $g$. The operator $E(\Delta_j)$ in (\ref{266})-(\ref{268})
is given by
\begin{equation}
E(\Delta_j)=\int\limits_{\Delta_j}\vert\phi_{q,p}\rangle d\Sigma_m(q,p)
\langle\phi_{q,p}\vert\label{269}
\end{equation}
where $\Delta_j\subset\Sigma^{\pm}_m=\sigma\times V^\pm_m$ with
$E(\Sigma^\pm_m)={\bf 1}^\pm$.
 
The POV property (\ref{266}) implies that for every normalized
state $\bfPsi^{(s)}\in\H^{(s)}_{\te}$ the expression
\begin{eqnarray}
P_\psi(\Delta_j)&=&\langle\bfPsi^{(s)}\mid E(\Delta_j)\bfPsi^{(s)}\rangle
\nonumber\\
&=&\int\limits_{\Delta_j\times\C_2}\vert\psi^{(s)}(q,p;\a)\vert^2
\delta({1\over mc} p_{A\dot A}\a^A\bar\a^{\dot A}-r)d\Sigma_m(p,q)
d\a d\bar\a
\label{270}
\end{eqnarray}
computed according to (\ref{264}) yields the probability of finding the
(free) particle (or antiparticle)
with mass $m$ and spin $s$ within the domain $\Delta_j\in
\B$ of stochastic phase space. (For a detailed discussion we refer to
Ref.~\cite{AP}.) The last equality in (\ref{270})
is obtained from (\ref{269})
together with (\ref{262}). The r.-h.\ side of (\ref{270})
finally yields, remembering (\ref{265}) for a normalized state,
\begin{equation}
P_\psi (\Delta_j)= N_s\int\limits_{\Delta_j}\sum\limits_{s_3}
\vert\psi^{(s)}_{s_3}(q,p)\vert^2 d\Sigma_m(q,p)\/,
\label{271}
\end{equation}
with
\begin{equation}
N_s =\left[~\int\limits_{\Sigma^{\pm}_m}\sum_{s_3}\vert\psi^{(s)}_{s_3}
(q,p)\vert^2 d\Sigma_m(q,p)\right]^{-1}\/.
\label{272}
\end{equation}
 
\vspace{1cm}

\section{Hilbert bundle over curved space-time with fiber $\Hs$}
In the previous section we have developed an internal space description
for spin leading, in the reduced case when Eqs.(\ref{17}) and (\ref{18})
are satisfied, to a quantum mechanical formulation for the kinematics of
free particles with mass $m$ and integer or half integer spin $s$ in
terms of scalar wave functions, $\psis$, realized on a one-particle
Hilbert space $\Hs$ carrying a unitary irreducible phase space
representation of the Poincar\'e group. From $\psis$ the usual
($2s+1$)-dimensional vector representation of the spin degrees of
freedom may easily be recovered. However, for many investigations it is
simpler to use the scalar functions $\psis$ directly together with their
Poincar\'e transformation rule (\ref{263})
and their invariant scalar product
(\ref{264}) characterizing the irreducible
resolution kernel Hilbert
space $\Hs$.
 
In this section we would like to describe free quantum particles of mass
$m$ and arbitrary but definite physical integer or half integer spin
$s$ in the presence of gravitation. We aim at a formulation in terms
of generalized wave functions defined on a  first quantized (i.e.\
one-particle or one-antiparticle) Hilbert bundel,
$\Hms$, raised over a curved
Riemann-Cartan space-time $U_4$ possessing a pseudo-Riemannian
metric and a metric compatible torsion. The standard fiber of $\Hms$
is the one-particle Hilbert space $\Hs$ constructed above carrying an
irreducible phase space representation, $U^{(s)}(b,\Lambda)$,
of the Poincar\'e group characterized in the Wigner sense by $[m,s]$.
The group action on $\Hms$ is given in terms of $U^{(s)}(b,\Lambda)=U^{(s)}
(g)$.
The basic properties of the Hilbert space $\Hs$ (for fixed physical
$[m,s]$) as resolution kernel Hilbert space with resolution
generator $\te$ are determined by the Hilbert space
$\H_{\te}=\H_{\te}^{(s=0)}$. The
spin description for free noninteracting particles of mass $m$
and nonvanishing spin $s$ in flat space
adds only an ``inessential complication''
described, as mentioned, by $(2s+1)$-component fields $\pss (q,p)$,
or -- more concisely and before performing a decomposition in terms of
homogeneous polynomials in the variable $\alpha$ -- described by the
scalar field $\psis$ depending on the internal spinor variable
$\a =(\a^1,\a^2)$, obeying, because of (\ref{257}) and (\ref{258}),
\begin{equation}
\hat D\psis = 2s\psis\/.
\label{31}
\end{equation}
 
The use of the $\C_2$-variables $\a,\bar\a$ for a general
description of spin prior to the
reduction to a particular value $s=0,\ha,1{3\over 2}\ldots$
for free (asymptotic) physical particles, which would be considered
in connection with a dynamical coupling of several
general spin fields in a theory
incorporating a dynamical role of spin, will be investigated
elsewhere. Here we want to concentrate on the formulation
of the {\it kinematics of free quantum particles} with definite mass
and arbitrary (but specified) physical spin, i.e.\ $[m,s]$ fixed,
in terms of {\it reduced} fields transforming as phase space
representations of the Poincar\'e group which is realized in the local
fibers of a bundle over space-time in the presence of gravitational
and possibly torsion fields, i.e.\ being given as sections on the
Hilbert bundle $\Hms$ over a Riemann-Cartan space-time $U_4$ defined
by
\begin{equation}
\Hms =\H (U_4,\F=\Hs\/,U^{(s)}(g))\/.
\label{32}
\end{equation}
The bundle $\Hms$ is associated to the Poincar\'e frame bundle over $U_4$
with structural group $G=ISO(3,1)\equiv\P$, i.e.
\begin{equation}
P =P (U_4,\P)\/,
\label{33}
\end{equation}
or, rather, to the corresponding spin
frame bundle $\bar P=\bar P (U_4,\bar\P)$
with structural group $\bar\P$ as far as the transformation of the
internal spinor variables $\alpha^1,\a^2$ are concerned. We add in
parenthesis that we shall assume a spin structure to exist on
space-time, i.e.\ we shall assume that $P$ is a trivial bundle
possessing global sections so that the homomorphism between $\bar\P$ and
$\P$ carries over to a corresponding homomorphism between the bundles
$\bar P$ and $P$.
 
A further bundle associated to $P$ is the one-particle phase space
bundle for zero spin
\begin{equation}
\tilde E=\tilde E_{s=0}=\tilde E(U_4,\tilde F=\Mpm, \P)
\label{34}
\end{equation}
with structural group $\P$ and standard fiber $\Mpm = M_4\times\Vpm$.
$\tilde E$ is a soldered bundle \cite{Drechsler} with first order
contact of the space-time base and the fiber over $x$, $\Mpm (x)$,
for each $x\in U_4$. The contact between base space and fiber is made
in $\tilde E$ through the local subspace $M_4\simeq T_x$ of $\Mpm (x)$.
The affine tangent bundle $T_A (U_4, F=M_4, \P)$, with the Minkowski
fiber viewed as an affine space with group of motion $\P$, is in a
natural manner submanifold of $\tilde E$. Disregarding spin,
an atlas on the bundle
$\tilde E$ provides the concrete kinematical
localization and momentum variables $(x;q,p)$ on which
the generalized wave functions, defined on $\Hms$, depend.
Here $x\in U_4$ is a classical space-time variable and $(q,p)\in\Mpm$
are local stochastic phase space variables corresponding to {\it mean}
position $q\in T_x(U_4)$ and {\it mean} momentum $p\in\Vpm$. (For a detailed
account of the geometro-stochastic
formalism and the basic fuzziness encoded in this description in
the fiber variables $(q,p)$ and the corresponding
resolution generator depenting on
a length parameter $l$ we refer to Refs.~\cite{prug1},\cite{prug2}
and \cite{AP}.)~ 

We now want to extend the geometro-stochastic (g-s)
description
for quantized one-particle
states on curved space-time to the bundle $\Hms$ for arbitrary
physical spin by including the spinor variables $\alpha^A$.
These latter variables will, however, at first {\it not} play the role of
stochastic variables for the description of spin. In this respect
$\alpha $ is different from the pair $(q,p)$ in the geometro-stochastic
formalism. From the later discussion of the quantum propagation
on $\Hms$ in the presence of gravitation, which is discussed in Sect. IV
below, we shall however find that the spin polarization of states
does finally also acquire a stochastic nature.
We thus first generalize (\ref{34}) to the classical
phase space bundle $\tilde E_s$ associated to $\bar \P$ for
single particles of type $[m,s]$, i.e. define
\begin{equation}
\tilde E_s=\tilde E_s(U_4,\tilde F_s=\Mpm\times S^2_{r=2s},\bar \P)~,
\label{34a}
\end{equation}
where the $SL(2,\C)$ part contained in $\bar\P$ acts on the
two-sphere, $S^2_{r=2s}$, as described in the introduction, and
with $\P$ (the homomorphic image of $\bar\P$) acting on $\Mpm$
in the usual way as in (\ref{34}), i.e. for $g=(b,\Lambda ) \in \P :
g(q,p)=(\Lambda q-b,\Lambda p)$. [Compare Eqs. (\ref{37}) and
(\ref{38}) below.]
The soldered bundle $\tilde E_s$ provides
the local variables $(q,p;\alpha )$
at the point $x$ of the base on which the generalized
geometro-stochastic wave functions for arbitrary physical mass and
spin, which are to be defined on $\Hms$, will depend.

Let us now first choose a (global) gauge on $P$ and denote it by
$\sigma_P(x)$. The corresponding coherent state base in the local fiber
$\Hs (x)$ of $\Hms$ is,
for any spin and for arbitrary $x\in U_4$, given by
\begin{equation}
\bar\sigma (x) : \quad \Fisi\longrightarrow\phi_{q,p}\/~,
\label{35}
\end{equation}
where we have denoted the map in $\Hms$ corresponding to
$\sigma_P$ by the symbol $\bar\sigma$, i.e. with a bar in order to
discriminate it from the space-like surfaces in $\Mpm (x)$
which are denoted by $\sigma (x)$ [compare (\ref{25}) and (\ref{220})].
The states denoted
by $\Fisi$ provide a {\it local coherent quantum frame basis} of
$\Hs (x)$ and yield a corresponding resolution of the identity
in the fiber over $x\in U_4$ in $\Hms$ independent of the value for
$s$~:
\begin{equation}
\int_{\Sipm}\vert\Fisi\rangle d\Sigma_m (q,p)\langle\Fisi\vert =
{\bf 1}^\pm_x~.
\label{36}
\end{equation}
Here $\Sipm$ denotes a subspace of $\Mpm (x)$ given by the direct
product of a space-like hypersurface $\si (x)$ in $T_x (U_4)$ and the 
hyperboloid $\Vpm (x)$. It is easy to show that a change of section
$\sigma_P (x)\longrightarrow\sigma'_P (x)$ on $P$
corresponds to an $x$-dependent Poincar\'e transformation for sections
on $\Hms$ in
the following manner (compare Ref. \cite{DT})
\begin{equation}
\bar\sigma'(x)=U^{(s)}(g(x))\bar\si (x)=\Us (b(x),\Lambda (x))\bar\si (x)
\label{37}
\end{equation}
with $g (x)=(b(x),\Lambda (x))$ acting on the local affine frame
$(\ua(x),e_j (x))$ in the gauge $\sigma_P(x)$ on $P$,
with $\ua(x)=-a^k(x)e_k(x)$
denoting its origin, yielding the local affine frame
$(\ua'(x),e'_k(x))$ in the gauge $\sigma'_P(x)$ on $P$ with
$\ua'(x)=-a'^k(x)e'_k(x)$ denoting the new origin.
The relations between the primed and unprimed frames are given in
components by
\begin{equation}
a'^k(x)=[\Lambda (x)]^k{}_j a^j(x)+b^j(x)\/;~~ e'_k(x)
=e_j(x)\left[\Lambda^{-1}(x)\right]^j{}_k~,
\label{38}
\end{equation}
where repeated local Lorentz indices are summed over $0,1,2,3$.
Moreover, $\Us (g(x))$ leaves (\ref{36}) invariant, where, in fact,
only the $s=0$ part is involved as mentioned.
 
A state of a particle of type $[m,s]$ in the Hilbert
bundle description on $\Hms$ is represented by a smooth section
\begin{equation}
x\longrightarrow \bPsi \in\Hs (x)\label{39}
\end{equation}
involving a state vector $\bPsi$ to be defined in each local
fiber of $\Hms$ above the base point $x\in U_4$. In analogy to
Eq.~(\ref{261}) the state vector $\bPsi$ may be decomposed with respect
to the local quantum frame basis $\Fisi$ according to
\begin{equation}
\bPsi =\int_{\Sipm}\psi^{(s)}_x(q,p;\a)~\Fisi d\Sigma_m(q,p)
\label{310}
\end{equation}
where
\begin{equation}
\psisx =\langle\Fisi\mid\bPsi\rangle_{\Sipm}\label{311}
\end{equation}
is the corresponding gauge dependent {\it generalized one-particle
geometro-stochastic (g-s) wave
function} defined on $\Hms$ which transforms under a change of section
(\ref{37}), i.e.\ under Poincar\'e gauge transformations, as
\begin{eqnarray}
\left[\psisx\right]'&=&\left(\Us (b(x),\Lambda (x))\psi^{(s)}_x\right)
    (q,p;\alpha)\nonumber\\
&=&\psi^{(s)}_x\left(\Lambda^{-1}(x)(q-b(x)),
\Lambda^{-1}(x)p,D(\Lambda^{-1} (x))\alpha\right)\/.
\label{312}
\end{eqnarray}
For ease of writing we have suppressed a lable $\bar\sigma (x)$
on $\psi^{(s)}_x$ in the equations above.
 
There is again a Poincar\'e gauge invariant scalar product defined
on $\Hms$ constructed as in Eq.~(\ref{264}), however now written with the
smooth $x$-dependent sections $\psi^{(s)}_{1,x}(q,p;\a)$
and $\psi^{(s)}_{2,x}(q,p;\a)$ defined on $\Hms$ and involving
invariant integration over $\Sigma^{\pm}_m\times\C_2$ at the
point $x$ of the space-time base.
 
The internal spinor variable $\a$ is now a local $SL(2,{\bf \C})$ gauge
variable comparable to the local kinematic Poincar\'e variables (i.e.\
the stochastic variables) $q$ and $p$. However, the reducing property
of the generalized wave functions $\psisx$ for a free particle of type
$[m,s]$ is independent upon $x$, i.e.\ Eq.~(\ref{31}) is valid for
$\psisx$ at any point $x$ on the base of $\Hms$ with the same relation
(\ref{31}) being satisfied also by the state vector $\bPsi$.
 
The covariant derivative of the generalized scalar g-s wave function
$\psisx$ is given by
\begin{equation}
D\psisx =[d+i\Gamma (x)]~\psisx
\label{313}
\end{equation}
with $d=\theta^k\partial_k$, where $\theta^k$; $k=0,1,2,3$ is a base of the
cotangent space $T^*_x (U_4)$ at $x\in U_4$ and, correspondingly,
$D=\theta^k D_k$. Furthermore, we denote by $(\tilde\theta^k (x),
\tilde\omega_{ij} (x))$ a connection on $P$
pulled back to the base under the map $\sigma_P$, where
\begin{equation}
\tilde\theta^k(x) =\theta^k +\nabla a^k(x)
\label{314}
\end{equation}
are the soldering forms defining the translational part of the
connection on the Poincar\'e bundle (\ref{33}) with $\nabla a^k(x)$
denoting the covariant derivative of the $k$-th component
of the translational part of the affine frame
field $(\ua(x), e_j (x))$ taken with respect
to the Lorentz part of the connection on (\ref{33}) given by
\begin{equation}
\tilde\omega_{ij} (x) = -\tilde\omega_{ji} (x)=\theta^k
\tilde\Gamma_{kij} (x)
\label{315}
\end{equation}
with coefficients $\tilde\Gamma_{kij} (x)$. In (\ref{313})
$\Gamma (x)$ may now be defined as (compare Refs. \cite{Drechsler}
and \cite{WD})
\begin{equation}
\Gamma (x) = -\tilde\theta^k(x)\tilde P_k +\ha \tilde\omega_{ij}(x)
\tilde M^{ij}+\ha \tilde\omega_{ij}(x)\tilde S^{ij}\/.
\label{316}
\end{equation}
Here $\tilde P_k, \tilde M_{ij}$ (with the indices $i,j$ lowered using
the Minkowski metric $\eta_{ik}$) are the generators of the phase space
representation $U^{(s=0)}(b,\Lambda )$ constructed  in terms of
differential operators in the variables $q_k$ and $p_k$, and
$\tilde S_{ij}$ are the corresponding operators of the
spin dependent part of the representation $U^{(s)}(b,\Lambda )$,
for $s\not= 0$, given as differential operators in the spin
space variables $\alpha^A,\bar\alpha^{\dot A}$ for definite $s$ which
are related to the generators of the $SL(2,{\bf \C})$
transformations $D(\Lambda )$ and $\bar D(\Lambda )$
in Eqs. (\ref{223a}) and (\ref{223b}) (compare \cite{BK}).
 
\vspace{1cm}

\section{Quantum Propagation on $\Hms$}
We are interested in the geometro-stochastic propagation, called
quantum propagation, on the bundle $\Hms$ of a generalized reduced
wave function (section) $\psisx$, defined in (\ref{311}) and (\ref{312}),
describing a single particle (or antiparticle) of definite physical
mass and spin $[m,s]$. The {\it phase space probability amplitude}
associated with the section $\psisx$ on $\Hms$ is given by
(compare the discussion presented
in \cite{DT}, \cite{ep1} and \cite{ep2}):
\begin{equation}
\psi^{(s)}(x,p;\a)=\psi^{(s)}_x(q=-a(x),p;\a)~.
\label{41}
\end{equation}
Here $q=-a(x)$ denotes the {\it point of contact} of $T_x(B)\subset
\M^\pm_m(x)$ with the space-time base $B=U_4$ 
on the bundle $\tilde E_s$ in any Poincar\'e gauge
on $P$. This point will be identified with the point $x$ of the base.
Furthermore, $p\in \Vpm (x)$ and $\a\in S^2_{r=2s}(x)$ in (\ref{41})
[compare Eq. (\ref{34a})].

The particle is described quantum mechanically and is considered to be
free except for influences of gravity described through the curvature
of the base of $\Hms$ which is treated as an external field.
No back reaction of the quantum particle onto the underlying
geometry is thus considered.
Clearly, the propagation on $\Hms$ conserves the
mass and spin value; hence the quantum propagator for $\psisx$ and the
associated probability amplitude (\ref{41}) has to commute with
$\hat D$ and with the Casimir operators $\hat P_\mu\hat P^\mu$ and
$\hat W_\mu\hat W^\mu$. Moreover, we found in Sect. III B above that
the stochastic phase space propagator $K^{(s)}_{\tilde\eta}(q',p';q,p)$
describing the propagation of a particle of spin $s$ in the local
fibers of $\Hms$ is independent of $s$.

On $\Hms$ the generalized one-particle wave function $\psisx$ should
be a solution of a second order wave equation,
\begin{equation}
(\Box_{\Hms}+\beta )\psisx=0~,
\label{42}
\end{equation}
where the invariant second order differential operator is
\begin{equation}
\Box_{\Hms}=g^{\mu\nu}\bar D_\mu D_\nu =\frac{1}{\sqrt {-g}}
D_\mu \sqrt {-g}~ g^{\mu\nu} D_\nu -
g^{\mu\nu}K_{\mu\nu}{}^\rho D_\rho  ~,
\label{43}
\end{equation}
with $D_\nu$ as defined in (\ref{313}), using $D_\nu=
\lambda^k_\nu (x)D_k$, where the $\lambda^k_\nu (x)$ are the vierbein
fields, and with $g_{\mu\nu}(x)
=\lambda^i_\mu (x)\lambda^k_\nu (x)\eta_{ik}$
being the covariant metric tensor
in the base of $\Hms$, and correspondingly for the contravariant
metric tensor $g^{\mu\nu}$. [Tensor components referring to a natural
basis, $\partial_\mu ~\!; \mu =0,1,2,3$, are labeled with Greek indices.]
$\bar D_\mu $ in (\ref{43})
is the Poincar\'e gauge covariant {\it and} $U_4$-covariant derivative,
and $K_{\mu\nu}{}^\rho$ denotes the torsion tensor. 
[For axial vector torsion,
considered below, the last term on the r.-h. side of Eq. (\ref{43})
is absent due to the antisymmetry of the $K_{\mu\nu\rho}$ in this case.]
In Eq. (\ref{42}) $\beta$ is an invariant of dimension $L^{-2}$ ($L$=
length), depending on $m$ and possibly on $s$, which characterizes the
wave motion on $\Hms$. To what extent $\beta$ contains
a $U_4-$ (or, in the
absence of torsion a $V_4-$) curvature invariant characterizing the
geometry of the base, as discussed in conformally invariant theories
\cite{P}, will not be made explicite here; compare, however, in this
context the work of Buchdahl for higher spin fields in
Riemannian spaces \cite{B} and the remarks made in Sect. V below.

We are aiming at a path integral-like solution of (\ref{42}), valid
for arbitrary integer or half integer spin, 
which is constructed in analogy
to Feynman's path integral representation of a nonrelativistic wave
function satisfying the Schr\"odinger equation \cite{RPF}.

In \cite{DT} a careful study was undertaken to show that a formula
conjectured by Prugove\v{c}ki
(compare Ref. \cite{ep1} as well as \cite{ep2})
for the quantum propagation on $\Hms$ for spinless
particles is indeed Poincar\'e gauge covariant (i.e. is Poincar\'e
gauge invariant except for endpoint transformations), it is
curvature and hence path dependent (i.e. is sensitive to the metric
curvature of the base), and yields the correct special relativistic
expression in the flat space limit. In this path integral-like
formula for the propagation on the Hilbert bundle one considers a
particular foliation of the space-time base into space-like
hypersurfaces $\sigma (\tau )$ with evolution parameter
$\tau$ and regards the surfaces $\sigma(\tau )$
through the point $x_0\in B$ for $\tau=\tau_0$ and $x=x_N\in B$
for $\tau=\tau_N$ after $N$ iterations, $n=1...N$. The
geometro-stochastic propagator for the probability amplitude
of a spinless massive particle is now defined by considering
{\it all} polygonal paths between $x_0$ and $x$ composed of
free-fall segments, i.e. constructed with geodesic arcs of the
underlying metric between points on two adjacent foiles
$\sigma(\tau_{n-1})$ and $\sigma(\tau_n)$ of the foliation.
One considers thus parallel transport on $\Hms$ between adjacent
points $x_{n-1}\in\sigma(\tau_{n-1})$ and $x_n\in\sigma(\tau_n)$
using different starting conditions regarding the stochastic
momentum variable in each step. The computation -- assumed to
apply to small space-time distances -- is unrestricted by
classical causality arguments, and integration in a Poincar\'e
gauge invariant manner over the {\it full} intermediate space-like
surfaces of the foliation is carried out like in relativistic
Feynman path integral formulations in Minkowski space \cite{F2,miu}, i.e.
without restricting the construction to the propagation along
broken paths composed of time-like segments only.
The quantum propagator for the amplitude, finally, results in the
limit $N\to\infty$, i.e. by making the geometro-stochastic
averaging involving broken polygonal paths finer and finer.

Before we continue our construction of a g-s propagator in the presence
of gravitation, let us inject here some brief remarks concerning
the so-called Einstein causality, observed
to hold for macroscopic distances
in space-time, arising in the present context as the result of the
superposition and destructive interference of probability amplitudes
originating from classically forbidden space-time regions.
The property of stochastic microcausality in the framework
of the stochastic phase space formulation of quantum mechanics
has been investigated
in detail by Greenwood and Prugove\v{c}ki \cite{GP} using the concept
of ``asymptotic causality" \cite{R}, i.e. the causal features arising
in the limit $\tau\to\infty$. Let us, however, first mention that
the stochastic phase space
propagator $K_{\tilde \eta_l} = K_{\tilde \eta}(q',p';q,p)$ in
flat space -- or, more exactly,
in $\Mpm = M_4\times\Vpm$ -- which was defined in Eq. (\ref{213}),
is, for small
stochastic smearing characterized by the fundamental length parameter
$l$ [see Sect. II A], indeed ``close" to the Feynman propagator
$i\Delta_F(q'-q)$ which is known to be nonzero for space-like
separation of the points $q'$ and $q$ 
(compare the discussion presented in \cite {GP}).
For finite (small) nonzero $l$ the stochastic phase space description
using generalized wave functions is formulated in terms of {\it spread
out quantum events} (at the scale of $l$) and, correspondingly, the
propagation of wave functions describing such events is only
``stochastically causal" and not deterministically causal in the strict
sense as in the yes-no manner realized in classical relativistic
physics with strictly zero influences on points outside the future
light cone of an idealized pointlike event localized at $q$.
In the stochastic setting
used here one has the result, obtained first for the flat space case
in \cite{GP}, that the probability for a particle of propagating
outside the future light cone of a certain point tends to zero
with $\tau\to\infty$. Thus no events violating Einstein causality do 
occur {\it in the infinite future} in this stochastic formalism. This
property has been called {\it asymptotic stochastic microcausality}.
Hence also in the presence of gravitation, i.e. for a curved base $B$,
the causal features of quantum propagation
will be stochastic in nature with Einstein causality being 
approached for infinitely separated (stochastic) events.

Continuing now our construction of a quantum propagator on $\Hms$
by means of parallel transport along broken paths composed of
geodesic segments, we may also consider that
the quantum particle would be measured by a certain localization device
with a given resolution in between the initial and final points
$x_0$ and $x$, respectively, and with their stochastic localization
given at these and at the intermediate points
in terms of the respective fiber variables.
The class of possible broken paths
composed of geodesic arcs would then have to be narrowed to a certain
corridor in the sense of Mensky \cite{mens}. We shall, however,
not discuss problems of this kind in the present paper and sum
over {\it all} intermediate broken trajectories. But even if the quantum
particle is not followed by continous measurement with a certain
resolution it is assumed that it keeps its identity with respect to its
mass and spin value. Hence one has to postulate, as mentioned above,
that the quantum propagator, which is path dependent for a curved base,
does commute with the Casimir operators defined in (\ref{11}) and
(\ref{12}) and with the spin operator $\hat D$.
While g-s propagation is required to conserve the spin value $s$
it will, however, affect the spin projection $s_3$ i.e. the
polarization of the state considered (see below).

In the works cited above the geodesic arcs and path dependences
were computed using the Levi-Civita connection embedded into the
Poincar\'e framework used here by putting the affine vector field
$a^k(x)$ in (\ref{314}) equal to zero and considering the pull back
of a connection on $P$ given by the one-forms 
$(\theta^k, \tilde\omega_{ij}(x))$. However, now $\tilde\omega_{ij}(x)$
may contain torsion effects for the base being a Riemann-Cartan
space-time $U_4$, i.e. 
$\tilde\omega_{ij}(x)=\bar\omega_{ij}(x)+\tau_{ij}(x)$, where
$\bar\omega_{ij}(x)$ is the purely metric part and $\tau_{ij}(x)$
is the torsion addition with $\tau_{ij}(x)=\theta^k K_{kij}(x)$.
For axial vector torsion [ i.e. for a completely antisymmetric
torsion tensor field $K_{kij}(x)$ ] no effects on the
geodesics would be possible. Since the role of torsion in this whole
context is not yet clear and since no source equations for the
determination of torsion -- supposed to be
induced in the underlying geometry
by a feed-back mechanism involving the quantum fields -- has been
discussed in this paper \cite{textb},
we shall assume that torsion is not
affecting the geodesics entering the definition of the quantum
propagator, i.e. we shall use $(\theta^k,\bar\omega_{ij}(x))$ as
connection one-forms in a particular gauge on $P$ and regard the
backgroung metric $g_{\mu\nu}(x)$ as determined by the solution
of Einstein's equations with given classical sources \cite{textc}.
The quantum
propagator for the $\psi $-field is then the ``free" g-s propagator
for a quantized test particle field under the influence of a classical
background metric and possibly also in the presence of an external
axial vector torsion field. Our aim here is to extend this 
kinematic description
to quantum particles of arbitrary integer or half integer spin $s$
by using the internal spin variables investigated in the previous
sections.

To this end we first quote the result for $s=0$ using (with a slight
change) the notation of Ref. \cite{DT} for the operator
$\underline K^{\bar\sigma}(x',q',p';x,q,p)$ of quantum propagation on
$\H_{[m,0]}$
(compare also Refs. \cite{ep1}
and \cite{ep2}):
\begin{eqnarray}
\underline K^{\bar\sigma}(x',q',p';x,q,p)=&&\lim_{N\to\infty}\int
K^{\bar\sigma}_{\gamma (x',x_{N-1})}
(x',q',p';x_{N-1},\hat q_{N-1},\hat p_{N-1})
\nonumber\\
&&\prod^1_{n=N-1}~
K^{\bar\sigma}_{\gamma (x_n,x_{n-1})}(x_n,\hat q_n,\hat p_n;
x_{n-1},\hat q_{n-1},\hat p_{n-1})~d\Sigma_m(x_n,\hat p_n)~.
\label{44}
\end{eqnarray}
Here we have replaced the complex variable $\zeta$ of \cite{DT}
by the pair $(q,p)$ and denoted the gauge by $\bar\sigma$ instead
of $s$ in order not to confuse it with the spin variable.
$K^{\bar\sigma}_{\gamma (x_n,x_{n-1})}$ represents the parallel
transport operator (in the gauge $\bar\sigma$) for parallel transport
from the point $x_{n-1}$ to the point $x_n$ along the geodesic arc
$\gamma ({x_n,x_{n-1})}$ in the base and $d\Sigma_m(x,\hat p)$ is
the ``contact point phase space measure" (compare \cite{DT}) given
by the measure defined in the local fiber of the bundle $\tilde E$,
introduced in (\ref{34}), restricted to the point of contact
of base space and fiber, i.e. evaluated for $q(x)=-a(x)\in T_x(B)$,
which is identified with the point $x$ of the space-time base $B$.
This procedure allows a Poincar\'e gauge invariant measure
to be associated with the leaves of a foliation of the space-time
base. Correspondingly, the intermediate fiber variables 
$(\hat q_n,\hat p_n)$, for $n=1,...(N-1)$, are given by
\begin{eqnarray}
\hat q_n &=& -a(x_n) \qquad {\rm identified~ with}~ x_n\in\sigma(\tau_n)
\subset B~,
\nonumber\\
\hat p_n &=& ~p(x_n) \qquad \in V^\pm_m(x_n)~.
\label{45}
\end{eqnarray}
In (\ref{44}), moreover, $x_0=x\in\sigma(\tau_0)$ with
$(\hat q_0,\hat p_0)=(q,p)\in\Mpm (x_0)$, and $x_N=x'\in\sigma(\tau_N)$
with $(\hat q_N,\hat p_N)=(q',p')\in\Mpm (x')$.

It was shown in \cite{DT} that with this interpretation of the
measure $d\Sigma_m(x_n,\hat p_n)=d\Sigma_m(-a(x_n),p(x_n))$
and integration over the intermediate variables $(x_n,p(x_n))\in
\sigma (\tau_n)\times V^\pm_m(x_n)$ for $n=1,...(N-1)$
the definition (\ref{44})
of a spin zero quantum propagator
$\underline K^{\bar\sigma}(x',q',p';x,q,p)$ is, indeed, Poincar\'e
gauge covariant (i.e. is Poincar\'e gauge invariant except for
transformations at the endpoints $x$ and $x'$ of the paths) and has
the correct flat space limit, where in a global Lorentz gauge
existing in that limit one can identify $K^{\bar\sigma}_\gamma$
(being path-independent in the flat space case) with the 
stochastic phase space
propagator $K_{\tilde\eta }$ defined in (\ref{213}). As a short-hand
notation we shall denote the domain of integration for
$\tau =\tau_n$, i.e. the hypersurface $\sigma (\tau_n)\times
V^\pm_m(x_n)$, by $\tilde\Sigma^\pm_m(x_n)$.

It is now straight forward to generalize the expression (\ref{44})
for spin zero to arbitrary physical values of spin by introducing
the internal spin variables $\a=\a(x)$ characterizing -- together
with the pair $(q,p)$ -- a point in the local fiber at $x\in B$
in the bundle $\tilde E_s$ defined in (\ref{34a}) and restrict
for the associated probability amplitude (\ref{41}) in the description
on $\Hms$ the $q$-value in the local tangent space
$T_x(B)\subset \M^\pm_m(x)$ to the point of contact, given in an
arbitray Poincar\'e gauge $\sigma_P$ on $P$, by $q=-a(x)$, and identify,
as mentioned, this point with the base point $x$.

However, before we discuss the generalization of Eq. (\ref{44})
let us remark that, since the parallel transport on $\Hms$ is
path dependent for a curved space-time base affecting for
$s\ne 0$ also the spin variable $\alpha$ [compare Eqs. (\ref{313})
and (\ref{316})], one would expect that starting with a generalized
decomposed wave function $\psi^{(s)}_{s_3,x}(q,p)$, obtained from
$\psi^{(s)}_x(q,p;\alpha )$ in analogy to (\ref{234}), having a sharp
spin projection value $s_3$ at a certain point $x\in \sigma (\tau_0)
\subset B$, there will appear -- as a result of the g-s propagation
involving {\it different} intermediate paths -- a spread in the
$s_3$-distribution of the spin projection value at the end point
of the paths. Hence, as the
result of the quantum propagation in the presence of gravitation
, i.e. for a metrically curved base, the spin projection $s_3$
of a certain state will become unsharp and develop a distribution
of values corresponding to a mixed state with unsharp (stochastic) spin
polarization. Such an effect of gravity on polarized spin states
should in principle be measurable at particle accelerators
provided it can be disantangeled from electromagnetic effects.

We now define the probability amplitude (\ref{41}) for definite
physical mass and spin at $x'\in B$ [corresponding to $q(x')=-a(x')$]
which results from the quantum propagation on $\Hms$ from the
amplitude prepared for $\tau =\tau_0$ on the hypersurface
$\sigma (\tau _0)\subset B$
[with $x\in \sigma(\tau_0)$ corresponding to $q(x)=-a(x)$] by
\begin{eqnarray}
\psi^{(s)}_{K_{x'x}}(x',p';\a')=
\int\limits_{\tilde\Sigma^\pm_m(x)\times\C_2(x)}
\underline K^{\bar\sigma ,(s)}{}(x',&&q',p',\a';x,q,p,\a)~
\psi^{(s)}(x,p;\a)
\nonumber\\
&&\times ~~\delta(\frac{1}{mc}p_{A\dot A}\a^A\bar\a^{\dot A}-r)
d\Sigma_m(x,p) d\a d\bar\a~,
\label{46}
\end{eqnarray}
where [compare (\ref{45})] $p'=p(x')\in\Vpm(x')$ and $p=p(x)\in\Vpm(x)$;
$q'=\hat q'=-a(x')$ [identified with $x'\in B$] and $q=\hat q=-a(x)$
[identified with $x\in B$]; $\a'=\a(x')\in\C_2(x')$ and $\a=\a(x)\in\C_2(x)$.
The $\delta$-function in (\ref{46}) guarantees, as in Eq. (\ref{264}),
that the integrations over the internal spin spaces, i.e. here
the $\C_2$-fibers at $x$ for all $x\in\sigma(\tau_0)$,
are restricted to the sphere $S^2_{r=2s}$ with
radius $r=2s$ corresponding to the spin value $s$ of the reduced
probability amplitude $\psi^{(s)}(x,p,\a)$ at $x$.
We denote the amplitude for spin $s$ of a state prepared at $x$
on the hypersurface $\sigma (\tau_0)$ and propagated to the point $x'$
on the hypersurface $\sigma (\tau ')$, for $\tau '=\tau_N$, by
$\psi^{(s)}_{K_{x'x}}(x',p';\a ')$. [Since we intend to construct
a solution of Eq. (\ref{42}) we may later drop the suffix $K_{x'x}$.]
Finally, $\underline K^{\bar\sigma , (s)}{}(x',q',p',\a';x,q,p,\a)$ in
(\ref{46}) is the quantum propagator for the probability amplitude
in the presence of spin given by the following expression:
\begin{eqnarray}
\underline K^{\bar\sigma ,(s)}{}(x',q',p',\a';&&x,q,p,\a)=
\lim_{N\to\infty}\int K^{\bar\sigma ,(s)}_{\gamma (x',x_{N-1})}
(x',q',p',\a';x_{N-1},\hat q_{N-1},\hat p_{N-1},\a_{N-1})
\nonumber\\
&&\times ~~
\prod^1_{n=N-1}~K^{\bar\sigma ,(s)}_{\gamma (x_n,x_{n-1})}
(x_n,\hat q_n,\hat p_n,\a_n;x_{n-1},\hat q_{n-1},\hat p_{n-1},\a_{n-1})
\nonumber\\
&&\qquad\times ~~
\delta (\frac{1}{mc}[\hat p_n]_{A\dot A}\a^A_n\bar\a^{\dot A}_n-r)
d\Sigma_m(x_n,\hat p_n) d\a_n d\bar\a_n~.
\label{47}
\end{eqnarray}
Eq. (\ref{47}) is analogous to (\ref{44}) and
the same notation is used for the variables $(\hat q,\hat p)$
as given in (\ref{45}). Moreover, $\a=\a_0=\a(x_0)\in \C_2(x_0)$
and $\a'=\a(x')\in\C_2(x')$ and analogously for the 
intermediate internal spin
variables $\a_n=\a(x_n)\in\C_2(x_n);~ n=1...(N-1)$, and their complex
conjugates. The intermediate integrations in (\ref{47}) run
over $\tilde\Sigma^\pm_m(x_n)\times\C_2(x_n);~ n=1...(N-1)$, i.e.
involve, due to the $\delta $-functions, the hypersurfaces
$\sigma (\tau_n)\times\Vpm(x_n)\times S^2_{r=2s}(x_n)$.

Clearly, $\underline K^{\bar\sigma ,(s)}$ is $s$-dependent and is
defined in a Poincar\'e gauge invariant manner except for endpoint
transformations of the variables $[q(x)=-a(x),p(x),\a(x)]$ at the
endpoints $x$ and $x'$ of the paths composed of geodesic arcs
in the base. Here the point $q(x)=-a(x)$, which
is identified with the point $x$ in the base, and analogously
the point $q(x')=-a(x')$, identified with $x'$,
remain unaffected by the gauge transformations.
The arguments proving the Poincar\'e gauge covariance 
of the expression for $\underline K^{\bar\sigma ,(s)}$ are the same as
those presented in \cite{DT} except for the additional internal spin
variables $\a$ appearing in (\ref{47}) together with
their Poincar\'e gauge invariant integrations
with the measure $d\a d\bar\a$
constrained by the $\delta$-functions. The propagator
$K^{\bar\sigma ,(s)}_{\gamma (x_n,x_{n-1})}$ in Eq. (\ref{47}), finally,
is the free fall propagator on $\Hms$ for the motion of a particle
along the geodesic arc
$\gamma (x_n,x_{n-1})$ from $x_{n-1}$ to $x_n$, which is obtained
as the solution of the differential equation
$D\psi^{(s)}_x(q,p,\a)=0$
[compare (\ref{313}) and (\ref{316})]
for parallel transport in $\Hms$ along $\gamma(x_n,x_{n-1})$
determining thus the propagator on $\Hms$
for the infinitesimal step from
$x_{n-1}\in\sigma (\tau_{n-1})$ to $x_n\in\sigma (\tau_n)$ of
the motion along the geodesic arc $\gamma(x_n,x_{n-1})$ in the base.

The path integral expression for $\psi^{(s)}_{K_{x'x}}(x',p';\a ,)$
constructed with ``free fall segments", i.e. with parallel shift along
geodesic arcs of the underlying metric, as defined by Eqs.
(\ref{46}) and (\ref{47}), obeying $D_\mu\psi^{(s)}=0$ for any
segment, is a solution of the second order wave equation (\ref{42})
provided the $\beta $-term in (\ref{42}) is zero by itself. This
requirement has the consequence that the mass and spin dependent
terms (if the latter is really there) must appear in such a way
that they compensate the curvature terms which might also be
present in the $\beta$-term. Hence a phenomonon which may be
called the ``Archimedes' principle"
must be at work setting mass and spin into correspondence
with an invariant curvature expression balancing thus these two
effects against one another: matter properties $[m,s]$,
on the one side, and
properties of the embedding geometry, on the other side.
The role played by torsion in this context is still unclear and
needs further study. However, it is apparent that torsion must play
the main part in this balancing since it is torsion which is
-- ultimately -- considered to be induced in the underlying geometry
as the ``footprint" of the quantum fields. In the present paper,
however, we investigate only the {\it kinematics} of supposedly
free (except for gravitation) spinning quantum particles and
regard torsion as an external field. We thus cannot
see this effect in detail without discussing field equations
for torsion at the same time.
Moreover, we should remember that torsion
has been severely restricted considering only axial vector type
(totally antisymmetric $K_{\mu\nu\rho}$).
\vspace{1cm}

\section{Discussion and Conclusion}
Following an idea proposed by Lur\c{c}at \cite{Lur} 
several decades ago, we discussed
in this paper the use of internal spin variables for a
quantum mechanical description of particles with real
positive mass $m$ and arbitrary integer or half-integer spin
$s=0,\ha,1,\frac{3}{2},2,\ldots$ in terms of {\it scalar functions}.
These generalized wave functions,
$\psi (q,p;\a,\bar\a)$, are defined over an extended phase space
$\Mpm\times\C_2$ in order to describe particles of
mass $m$ but arbitrary unspecified
spin, where $\Mpm=M_4\times\Vpm$ is the one-particle (+) or
one-antiparticle $(-)$ phase space with $q\in M_4$ and $p\in\Vpm$,
$p^2=m^2c^2$, $\pm\! =\! {\rm sign}~\! p_0$, and $\a$
denotes a point in the internal
spin space being a homogenous space of the Lorentz group of the type
$\bar S=SL(2,\C)/\bar H$ characterized by the subgroup $\bar H$
of $SL(2,\C)$ as explained in the introduction.
Following Bacry and Kihlberg \cite{BK} in choosing the
lowest dimensional internal spin space possessing a measure and being
capable of representing integer as well as half integer spins, we
used a four-dimensional internal spin space parametrized in terms
of spinor variables $\a\in\C_2$ with $\bar\a$ denoting the
corresponding complex conjugate spinor (dotted spinor) varying in the
complex conjugate spin space.

The one-particle wave function $\psi^{(s)}(q,p;\a )$
representing a particle ($p_0>0$) or
anti-particle ($p_0<0$) of definite mass and {\it fixed} integer or
half-integer spin, $[m,s]$, are then obtained by requiring that the
Eqs. (\ref{11}), (\ref{17}) and (\ref{18}) be satisfied yielding
thereby -- as far as the variables $(q,p)$ are concerned [playing
the role of stochastic variables in this context] -- an irreducible
element of a resolution kernel Hilbert space with resolution
generator $\tilde\eta=\tilde\eta_l$, and -- as far as the spin variables
$(\a,\bar\a)$ are concerned -- an irreducible element depending on $\a$
only (without dependence on $\bar\a$) with $\a$ varying on a two-sphere
$S^2_{r=2s}$ implying, as a consequence of demanding the homogeneity
condition (\ref{17}), that $\psi^{(s)}(q,p;\a)$ is a homogenous
polynomial of degree $2s$ in the undotted spinor variables $\a^A; A=1,2$
with no dependence on the dotted spinor variables $\bar \a^{\dot A};
\dot A=\dot 1,\dot 2$. The function $\psi^{(s)}(q,p;\a)$ may be
decomposed with respect to a basis transforming under the representation
$D^{(s,0)}$ of $SL(2,\C)$ to yield the familiar $(2s+1)$-dimensional
vector representation of spin, $\psi^{(s)}_{s_3}(q,p);~ 
s_3=-s\ldots +\!s$,
leading thus, ultimately, to a stochastic phase space description for
free particles (or antiparticles) of definite mass and physical
spin, $[m,s]$, transforming irreducibly under the Poincar\'e group
[compare Eqs. (\ref{240}), (\ref{257}) and (\ref{263})] and possessing
stochastic localization properties as far as the variables $(q,p)$ are
concerned. These functions are elements of the Hilbert space $\Hs=
L^2(\Sigma^\pm_m)\times K_s$ carrying an irreducible representation
of the covering group of the Poincar\'e group, $\bar\P$, characterized
by $m$ and $s$. This one-particle (or one-antiparticle) stochastic
phase space formulation for free particles of type $[m,s]$ in flat
space was then generalized to a formulation on a Hilbert bundle
$\Hms$ with fiber $\Hs$, being associated to the Poincar\'e spin
frame bundle over a curved Riemann-Cartan space-time base
possessing metric and torsion (the latter restricted to axial vector
type) both treated as external fields.
The aim was to derive a path integral-type
expression for the geometro-stochastic propagation of fields
for arbitrary physical mass and spin defined on the soldered Hilbert
bundle $\Hms$ constructed over a curved classical space-time base.
Our result for the quantum propagation is given by Eqs. (\ref{46})
and (\ref{47}) containing besides the Poincar\'e gauge invariant
intagrations over the intermediate phase space variables $(x,p)$
with measure
$d\Sigma_m(x,p(x))=d\Sigma_m(q(x)=-a(x),p(x))$ the Poincar\'e
gauge invariant
integrations over the intermediate internal spin variables with measure
$d\a d\bar\a$. The integrations over the spin variables are constrained
by delta functions -- coupling momentum and spin variables --
restricting the integrations to a particular
``spin shell", $S^2_{r=2s}$, for a particle with spin $s$ in
analogy to the momentum integrations restricted to the ``mass shell",
$\Vpm$, for a particle of mass $m$.
In this framework the stochastic localization properties
as well as the spin properties are described by means of the local
fibers of the bundle $\Hms$.

It was pointed out in Sect. IV that, although in the beginning
only $(q,p)$ were stochastic variables while the spin variables
$(\a ,\bar\a)$ were not of this type, with $s$ and $s_3$ taking
sharp values for a certain quantum state describing a particle of
spin $s$ and spin projection (polarization) $s_3$, the quantum 
propagation of such states on a curved space-time background (i.e.
in the presence of gravitation) leads, according to Eqs. (\ref{46})
and (\ref{47}), to a stochastic nature also for the polarization
of the states $\psi^{(s)}_{K_{x'x}}$ at $x'$ when decomposed at
that point, i.e. leads to a stochastic nature of the spin projection
$s_3$. To investigate this result, let us use an analyzing (or detection)
field at $x'\in\sigma (\tau ')\subset B$ and denote it by
$\psi^{(s)}_D(x',p';\a ')$ corresponding to a certain sharp
$s_3$-value when decomposed (representing, say, a state filtered by
a Stern-Gerlach magnet which is oriented in a certain way).
Then the invariant matrix element measuring the overlap at $x'$
between the originally prepared sharp spin state on $\sigma(\tau )$
with, say, $s=s_3$, propagated to $x'$, and a sharp detection field
with various settings of $s_3$ at $x'$ is given by
\begin{equation}
\langle\psi^{(s)}_D(x',p';\a ')\mid\psi^{(s)}_{K_{x'x}}(x',p';\a ')
\rangle_{\Sigma^\pm_m(x')\times\C_2(x')} ~.
\label{51}
\end{equation}
Sloppily stated the measuring procedure is the following: Produce a
pure measuring or detection state at $x'$ on the hypersurface
$\sigma (\tau ')$ and let it interfere with the state propagated
to $x'$ from all $x$ on the hypersurface $\sigma (\tau )$.
The analyzing or detection field $\psi^{(s)}_D(x',p;\a ')$
may then be varied with respect to the $s_3$-polarization
involved and the $s_3$-spectrum of $\psi^{(s)}_{K_{x'x}}(x',p';\a ')$
be measured in this way in order to determine what effect the quantum
propagation on a curved base had on an originally pure spin state,
i.e. how gravitation affected the propagation of the pure state
prepared on the hypersurface $\sigma (\tau)$.

As was also discussed in the previous section, the quantum propagation on
$\Hms$ is not causal in a classical sense (Einstein causality) but
is ``stochastically causal" \cite{GP}. Furthermore, we remarked at
the end of the section that the path integral representation
of the probability amplitude associated with a section on $\Hms$
(a generalized geometro-stochastic one-particle or one-antiparticle
wave function) satisfies a certain invariant second order wave equation
on $\Hms$ with certain restrictions imposed on the curvature
invariants appearing in the term denoted by $\beta$: The $\beta$-term
in (\ref{42}) had to vanish by itself compensating thus mass and
possibly spin dependent terms against invariant curvature
contributions. This we called ``Archimedes' principle" expecting that
torsion plays the dominant role in it. Let us point out again that
torsion was severely restricted in this context by
allowing only axial vector torsion from the beginning.

We did not discuss coupled first order spinor equations for arbitrary
spin which, historically, are known to develop inconsistencies for
$s\ge\frac{3}{2}$ when the minimal electromagnetic coupling is
introduced \cite{FP} or when these equations are generalized from flat
to curved space-time (possibly with torsion).
To make these equations consistent usually various supplementary 
conditions have to be imposed, i.e. auxiliary fields must be introduced
which render the resulting expressions very complicated and difficult
to handle. Instead we give here an analytic description of spin in
terms of internal variables for scalar functions based, as mentioned,
on Lur\c{c}at's
idea that spin should be described in terms of variables defined
on a homogeneous space of the underlying kinematic symmetry
group i.e. the Poincar\'e group. In fact, also in curved space-time the
Poincar\'e group may be considered, namely as gauge or structural group
of a soldered bundle raised over space-time, acting there on the
local phase space fiber variables (used there to describe
the [stochastic]
localization of quantized states) as well as on the internal spin
variables. It is thus indeed possible to give a general formulation
of one-particle states for arbitrary mass and spin in terms of scalar
functions and project out the conventional $(2s+1)$-component
vector states whenever necessary. However, for the understanding of
the quantum propagation of such fields in the presence of gravitation
it may be preferable to use the original generalized scalar
wave functions.

In concluding we remark that the stochastic phase space description
for single free relaticistic particles of arbitrary spin on a Hilbert
bundle over curved space-time $B$, which we studied in this papaer, may
easily be generalized to the many-particle case by considering
Fock bundles over space-time for particles of type $[m,s]$. The
standard fiber of these bundles are tensor products of one-particle
and one-antiparticle Hilbert spaces $\H^{(s)(+)}_{\te }$ and
$\H^{(s)(-)}_{\te }$, respectively, 
where $\H^{(s)}_{\te } = \H^{(s)(+)}_{\te }
\oplus \H^{(s)(-)}_{\te }$ with $(\pm )$ denoting the sign of the energy.
In order to be in accord with the Pauli principle one has to
introduce Fock bundles $\F_{[m,s]}$ possessing a fiber which is
a sum of {\it symmetrized} products for integer spin (bosonic case),
i.e.
\begin{equation}
\F^{(s)}_{sym} = \Bigl(\sum^\infty_{n=1}\otimes^n_{sym}\H^{(s)(+)}_{\te }
\Bigr)\otimes\Bigl(\sum^\infty_{n'=1}\otimes^{n'}_{sym}\H^{(s)(-)}_{\te }
\Bigr)\qquad {\rm for}~s=0,1,2\dots ~,
\label{52}
\end{equation}
and which is a sum of {\it antisymmetrized} products for half integer
spin (fermionic case), i.e.
\begin{equation}
\F^{(s)}_{anti} = \Bigl(\sum^\infty_{n=1}\otimes^n_{anti}\H^{(s)(+)}_{\te }
\Bigr)\otimes\Bigl(\sum^\infty_{n'=1}\otimes^{n'}_{anti}\H^{(s)(-)}_{\te }
\Bigr)\qquad {\rm for}~s=\ha,\frac{3}{2},\frac{5}{2}\dots ~.
\label{53}
\end{equation}
The Fock bundle of type $[m,s]$ associated to $\bar P$ [compare
Eq. (\ref{33})] is thus
\begin{equation}
\F_{[m,s]} = \F(B=U_4,~\F^{(s)}_{sym/anti},~U^{(s)}(g))
\label{54}
\end{equation}
with standard fiber (\ref{52}) for $2s$ being even, and with standard
fiber (\ref{53}) for $2s$ being odd. It is implied here that there
exists a local vacuum state $\mid \!{\bf O}\rangle_x$,
for every $x\in B$, which
is invariant under changes of sections on $\F_{[m,s]}$ provided by
Poincar\'e gauge transformations.

\vspace{1cm}

\section*{Acknowledgement}
I thank Drs. A. D. Popov and P. A. Tuckey for discussions
during the early stages of this work.

%
%
 
%
%
 

\begin{references}
\bibitem{Weyl} H. Weyl, {\sl Gruppentheorie und Quantenmechanik}
        2nd Edition, Hirzel Leipzig, 1931.
\bibitem{Wigner} E.P. Wigner, {\sl Group Theory and its Application to
        the Quantum Mechanics of Atomic Spectra},
        Academic Press, New York,
        1959.
\bibitem{Ed} A.R. Edmonds, {\sl Angular Momentum in Quantum Mechanics},
        Princeton University Press, Princeton, New Jersey, 1957.
\bibitem{Wig} E.P. Wigner, {\sl Ann. Math.} {\bf 40}, 149 (1939).
\bibitem{text} Here $\psi(x)~$[$\bar\psi(x)$] or
      $\psi_N(x)$~[$\bar\psi_N (x)$]
      represent, as usual, the four-component
      Dirac spinor fields for the electron or the
      nucleon isospinor field [and their adjoint], while $A^\mu(x)$
      is the spin~1 photon field and $\vec\phi (x)$ is the pseudoscalar
      isovector pion field. $\g_\mu$ and $\g_5$ denote, as usual,
      the Dirac
      $\g$-matrices, and $\vec\tau=(\tau_1,\tau_2,\tau_3)$ are the
      Pauli matrices for the isospin degrees of freedom;
      $e$ and $g$ are the respective coupling
      constants.
\bibitem{Lur} F. Lur\c{c}at, Physics {\bf 1}, 95 (1964).
\bibitem{BN} H. Bacry and J. Nuyts, {\sl Phys. Rev.} {\bf 157},
     1471 (1967).
\bibitem{Finkel} D. Finkelstein, {\sl Phys. Rev.} {\bf 100}, 924 (1955).
\bibitem{BK} H. Bacry and A. Kihlberg, {\sl J. Math. Phys.} {\bf 10},
       2132 (1969).
\bibitem{space} In Ref. \cite{BK}
      the homogeneous space $\bar S$ is denoted by
      $\bar H$.
\bibitem{alpha} S. Schlieder, in {\sl Quanten und Felder},
     Ed.~H.P. D\"urr,
     Fr. Vieweg, Braunschweig 1971.
\bibitem{beta} G.C. Hegerfeldt, {\sl Phys. Rev.} {\bf D10}, 3320 (1974);\\
     {\sl Phys. Rev. Lett.}, {\bf 54}, 2395 (1985) and\\
     {\sl Nucl. Phys. B} (Proc. Suppl.) {\bf 6} 231 (1989).
\bibitem{prug1} E. Prugove\v{c}ki, {\sl Stochastic Quantum Mechanics and
      Quantum Spacetime}, D.~Reidel, Dordrecht 1984.
\bibitem{prug2} E. Prugove\v{c}ki, {\sl Quantum Geometry}, Kluwer, Dordrecht
      1992.
\bibitem{Ali} Compare also Ali and Prugove\v{c}ki \cite{AP}, and
     Ali \cite{Ali2}.
\bibitem{AP} S.T. Ali and E. Prugove\v{c}ki, {\sl Acta Applicandae Math.}
     {\bf 6}, 1 (1986).
\bibitem{Ali2} S.T. Ali, {\sl Stochastic localization,
      quantum mechanics on
      phase space and quantum spacetime}, La Rivista del Nuovo Cimento
      {\bf 8}. 1-128 (1985).
\bibitem{Wood} N. Woodhouse,
      {\sl Geometric Quantization}, Clarendon Press,
      Oxford 1980.
\bibitem{Kir} A.A. Kirillov, {\sl Elements of the Theory of
    Representations}, Springer Verlag, \\Heidelberg 1976.
\bibitem{Evans} M. Evans, F. G\"ursey and V. Ogievetsky,
     {\sl Phys. Rev.} {\bf D47} 3496 (1993).
\bibitem{Popov} I thank Dr. A. Popov for a discussion on this point.
\bibitem{AliP} S.T. Ali and E. Prugove\v{c}ki,
     {\sl Acta Applicandae Math.} {\bf 6},
     47 (1986).
\bibitem{EPW} E.P. Wigner, {\sl Phys. Rev.} {\bf 40}, 749 (1932).
\bibitem{texta} For ease of writing we shall continue to denote the
     internal spin space simply by $\C_2$ although before the
     reduction to a definite spin value $s=\ha,1,\frac{3}{2},2\dots$ (see
     below) the spin space would actually be $\C_2\times\bar\C_2$,
     where $\bar\C_2$ represents the complex conjugate space of the
     dotted spinors with variables
     $\bar\a=(\bar\a^{\dot A};\dot A=\dot 1,\dot 2)$.
\bibitem{Carr} P.A. Carruthers, {\sl Spin and Isospin in
Particle Physics},
      Gordon and Breach, \\New York 1971.
\bibitem{Drechsler} W. Drechsler, {\sl Fortschr.\ Phys.}
        {\bf 38}, 63 (1990).
\bibitem{DT} W. Drechsler and P. A. Tuckey, {\sl Class. Quantum Grav.}
        {\bf 13}, 611 (1996).
\bibitem{WD} W. Drechsler, {\sl Ann.\ Inst.\ Henri Poincar\'e}
        {\bf 37}, 155 (1982) and {\sl Fortschr.\ Phys.} {\bf 32}, 
        449 (1984).
\bibitem{ep1} E. Prugove\v{c}ki, {\sl Principles of Quantum General
        Relativity}, Singapore 1995, World Scientific.
\bibitem{ep2} E. Prugove\v{c}ki, {\sl Class. Quantum Grav.} {\bf 11},
        1981 (1994), and {\bf 13}, 1007 (1996).
\bibitem{P} R. Penrose, {\sl Proc. Roy. Soc. (London)} {\bf 284},
        159 (1965).
\bibitem{B} H. A. Buchdahl, {\sl J. Phys. A: Math. Gen.} {\bf 15},
        1 (1982).
\bibitem{RPF} R. P. Feynman, {\sl Rev. Mod. Phys.} {\bf 20}, 367 (1948).
\bibitem{F2} R. P. Feynman, {\sl Phys. Rev.} {\bf 80}, 440 (1950),
             Appendix A.
\bibitem{miu} T. Miura, {\sl Progr. Theor. Phys.} {\bf 61}, 1521 (1979).
\bibitem{GP} D. P. Greenwood and E. Prugove\v{c}ki, {\sl Found. Phys.}
        {\bf 14}, 883 (1984).
\bibitem{R} S. N. M. Ruijsenaars, {\sl Ann. Phys. (New York)}
        {\bf 137}, 33 (1981).
\bibitem{mens} M. B. Mensky, {\sl Continous Quantum Measurement and
        Path Integrals}, Bristol 1993, IOP-Publishing.
\bibitem{textb} For a discussion of torsion effects in a theory
        based on the de Sitter group $SO(4,1)$, related to the
        Poincar\'e group treated here by group contraction with
        respect to the common subgroup $SO(3,1)$, we refer 
        to \cite{WE} and the literature quoted there.
\bibitem{WE} W. Drechsler, {\sl Quantized Matter in a de Sitter
        Gauge Theory with Classical Metric and Axial Torsion},
        Proceedings of the XIVth Course of the International School
        of Cosmology and Gravitation held at Erice, Italy, May 11-19,
        1995. World Sientific to be published.
\bibitem{textc} Clearly, one could use in the construction of
        the quantum propagator also the autoparallels of an underlying
        {\it general} Riemann-Cartan space-time base $U_4$ if the torsion
        tensor were known from other sources.
\bibitem{FP} M. Fierz and W. Pauli, {\sl Proc. Roy. Soc. (London)} 
        {\bf A 173}, 211 (1939).

\end{references}
\end{document}